\documentclass[journal]{IEEEtran}
\usepackage{cite}
\usepackage[colorlinks]{hyperref}
\usepackage{subfigure}
\usepackage{color}
\usepackage{graphicx}
\usepackage{booktabs}
\usepackage{amsfonts,amssymb}
\ifCLASSINFOpdf
\else
\fi
\begin{document}
\title{A Unified End-to-End Framework for Efficient Deep Image Compression}

\author{Jiaheng Liu,
        Guo Lu,
        Zhihao Hu,
        Dong Xu,~\IEEEmembership{Fellow,~IEEE}% <-this % stops a space
\thanks{Jiaheng Liu is with the School of Computer Science and Engineering, Beihang University, Beijing, China (e-mail: liujiaheng@buaa.edu.cn)}% <-this % stops a space
\thanks{Guo Lu is with Department of Electronic Engineering, Shanghai Jiao Tong
University, 200240 Shanghai, China (e-mail: luguo2014@sjtu.edu.cn)}% <-this % stops a space
\thanks{Zhihao Hu is with the School of Software, Beihang University, Beijing, China (e-mail: huzhihao@buaa.edu.cn)}% <-this % stops a space
  \thanks{Dong Xu is with the School of Electrical and Information Engineering, The University of Sydney, NSW 2006, Australia (e-mail: dong.xu@sydney.edu.au)}}% <-this % stops a space
% \thanks{Manuscript received April 19, 2005; revised August 26, 2015.}}

% \markboth{Journal of \LaTeX\ Class Files,~Vol.~14, No.~8, August~2015}%
% {Shell \MakeLowercase{\textit{et al.}}: Bare Demo of IEEEtran.cls for IEEE Journals}

\maketitle

\begin{abstract}
  Image compression is a widely used technique to reduce the spatial redundancy in images. 
  Recently, learning based image compression has achieved significant progress by using the powerful representation ability from neural networks. 
  However, the current state-of-the-art learning based image compression methods suffer from the huge computational cost, which limits their capacity for practical applications. 
  In this paper, we propose a unified framework called Efficient Deep Image Compression (EDIC) based on three new technologies, 
  including a channel attention module, a Gaussian mixture model and a decoder-side enhancement module. 
  Specifically, we design an auto-encoder style network for learning based image compression. 
  To improve the coding efficiency, we exploit the channel relationship between latent representations by using the channel attention module. 
  Besides, the Gaussian mixture model is introduced for the entropy model and improves the accuracy for bitrate estimation.
  Furthermore, we introduce the decoder-side enhancement module to further improve image compression performance.
  Our EDIC method can also be readily incorporated with the Deep Video Compression (DVC) framework~\cite{lu2019dvc} to further improve the video compression performance. 
  Simultaneously, our EDIC method boosts the coding performance significantly while bringing slightly increased computational cost.
  More importantly, experimental results demonstrate that the proposed approach outperforms the current state-of-the-art image compression methods and is up to more than 150 times faster in terms of decoding speed when compared with Minnen's method~\cite{minnen2018joint}.
  %  with input resolution of $768\times512$.
  The proposed framework also successfully improves the performance of the recent deep video compression system DVC~\cite{lu2019dvc}.
 Our code will be released at \url{https://github.com/liujiaheng/compression}.
  \end{abstract}
  

  \IEEEpeerreviewmaketitle

  \section{Introduction}
  Image compression aims to reduce the spatial redundancy in images and is widely used to save the bandwidth and storage sizes in lots of applications. 
  Traditional image compression methods ~\cite{wallace1992jpeg,skodras2001jpeg,WebP,BPG} rely on hand-crafted techniques to improve the compression efficiency.
  For example, JPEG \cite{wallace1992jpeg} uses the discrete cosine transform (DCT) to convert the images from the pixel domain to the frequency domain for high compression efficiency. However, the traditional compression methods cannot be optimized by using large-scale training, which may limit their performance.
  
  Recently, learning based image and video compression methods~\cite{toderici2015variable,toderici2017full,balle2016end, balle2018variational, theis2017lossy, agustsson2017soft,li2017learning, rippel2017real,mentzer2018conditional,agustsson2018generative,johnston2017improved,minnen2018joint,lee2018context,Wu_2018_ECCV,lu2019dvc} attract more and more attention.
  Ball{\'{e}} \textit{et al.} \cite{balle2016end} propose an end-to-end optimized image compression approach by using the convolutional neural network (CNN) based auto-encoder. To further improve the compression efficiency, Minnen \textit{et al.}~\cite{minnen2018joint} employs the auto-regressive prior information to obtain accurate entropy model and achieve comparable or even better performance than the traditional codec \cite{BPG}.
  
  Although the current state-of-the-art learning based methods \cite{lee2018context,minnen2018joint} improve the compression performance, they also increase the computational cost significantly.
When compared with the previous learning approaches \cite{balle2016end,balle2018variational}, the current state-of-the-art methods \cite{lee2018context,minnen2018joint} exploit the spatial redundancy in the latent feature space by using auto-regressive prior information. 
  Therefore, the decoding procedure in \cite{lee2018context,minnen2018joint} is performed sequentially for each pixel, while the previous approaches \cite{balle2016end,balle2018variational} can reconstruct all the pixels through convolution layers in a parallel manner.
  As shown in Table~\ref{tab:speed}, the average GPU decoding time for images with the resolution of $768 \times 512$ using Ball{\'{e}}'s method~\cite{balle2018variational} is \textbf{0.013 seconds} while the corresponding decoding time using Minnen's method \cite{minnen2018joint} is \textbf{2.426 seconds}. 
  % ICLR18 BD-RATE:  0.2986506007035754
  % NIPS18 BD-RATE:  0.5313589116428905
  % NO context+Ours BD-RATE:  0.5334900389336015
  \begin{table}
    \centering
    \caption{Decoding Time and BDBR improvement over JPEG2000~\cite{skodras2001jpeg} of different methods on the Kodak~\cite{Kodak} image dataset.
    The Deoding time is evaluating using one RTX 2080Ti.
    The full name of BDBR is ``Bjontegaard delta bitrate'',
    which refers to the bitrate relative percentage of reduction under the same PSNR.}
    \begin{tabular}{lll}  
    \toprule
    Methods  & Decoding Time & BDBR\\
    \midrule
    Ball{\'{e}}'s~\cite{balle2018variational}  & 0.013s& $29.87\%$\\
    Minnen's~\cite{minnen2018joint} & 2.426s  &  $53.14\%$\\ 
    EDIC(Ours) & 0.016s &$53.35\%$ \\
    \bottomrule
    \end{tabular}
    % \caption{Decoding Time and BD-BR improvement over JPEG2000~\cite{skodras2001jpeg} of different methods. The Deoding time is benchmarked on RTX 2080Ti.}
    \label{tab:speed}
    \end{table}
  In this paper, we ask the question: \textbf{Is it possible to improve the compression efficiency without significantly increasing the computation time?}
  To address this issue, we propose a unified framework named as Efficient Deep Image Compression (EDIC),
  which consists of three new components, including the channel attention module, the Gaussian mixture model and the decoder-side enhancement module. 
  Specifically, we utilize an auto-encoder style network for building the image compression framework.
  To further improve the compression performance, we also exploit the channel relationship in latent features at the encoder side and use an effective channel attention module to enhance the corresponding representation power. 
  More importantly, instead of using the single Gaussian model for entropy estimation like \cite{balle2016end,balle2018variational,minnen2018joint,lee2018context}, 
  we propose to use Gaussian mixture model (GMM) for more accurate entropy estimation.
  Besides, we introduce the decoder-side enhancement module to reduce the compression artifacts.
  The channel attention technique, the Gaussian mixture model and the decoder-side enhancement module are seamlessly combined,
  which leads to much better image compression performance with only slightly increased computational cost when compared with auto-regressive prior technique in \cite{minnen2018joint,lee2018context}.
  Experimental results demonstrate that the proposed image compression approach achieves comparable compression performance when compared with the current state-of-the-art approach \cite{minnen2018joint},
  while the decoding speed of our method is over \textbf{150} times faster than \cite{minnen2018joint} for images with the resolution of $768\times512$.
  Our method can be readily used for video compression and also achieves promising results for video compression.
  
  The contributions of this paper are summarized in the following aspects. First, to the best of our knowledge, we are the first to introduce the channel attention technique to improve image compression efficiency. Second, the Gaussian mixture model is introduced to model the distribution of the latent representation in a more accurate way. 
  Third, we additionally apply the decoder-side enhancement module to further improve image compression performance.
  Fourth, our proposed EDIC framework achieves the state-of-the-art image compression performance 
  while significantly reducing the decoding time when compared to Minnen's method~\cite{minnen2018joint}. 
  Fifth, the proposed framework is general and 
  also improve the performance of the recent learning based video compression system~\cite{lu2019dvc}.
  
  % It should be mentioned that there are two concurrent works, which propose similar ideas to boost the performance of image compression.
  % However, the publish date of our preprint version is close to these two works' preprint versions on arxiv.org.
  % Therefore, our work is considered as an independent work and our contributions are sufficient.
  Similar ideas were also proposed to boost the image compression performance in two recent works~\cite{cheng2020learned, lee2019hybrid}. 
  However, we would like to highlight that our work is a concurrent work as both works~\cite{cheng2020learned, lee2019hybrid} as the dates that the three works appeared in arxiv.org are very close to each other. 
  Moreover, our work is different with both works~\cite{cheng2020learned, lee2019hybrid} in the following three aspects. 
  First,  both works~\cite{cheng2020learned, lee2019hybrid} utilize the context information based on the work~\cite{minnen2018joint},
  which are very slow as the method in~\cite{minnen2018joint}. 
  In contrast, our work builds upon the method in~\cite{balle2018variational} instead of the algorithm in~\cite{minnen2018joint},
  thus our work is much faster than both works ~\cite{cheng2020learned, lee2019hybrid}.
  Second, our proposed attention module is to exploit the channel-wise relationships of the latent representations,
  while Cheng \textit{et al.}~\cite{cheng2020learned} introduce spatial attention scheme for image compression.
  Third, we also use our newly proposed technologies (i.e., Gaussian Mixture model, Channel attention scheme and Decoder-side enhancement method) for video compression and achieve promising results on the benchmark datasets,
  which are not discussed in~\cite{cheng2020learned, lee2019hybrid}.
  % {\color{red}Besides, Cheng \textit{et al.}~\cite{cheng2020learned} proposed to apply Gaussian mixture model and attention mechanisms to enhance the capacity of image compression.
  %     Lee \textit{et al.}~\cite{lee2019hybrid} also introduce Gaussian Mixture model and post-processing module to boost the performance of image compression.
  %     The methods proposed by~\cite{cheng2020learned, lee2019hybrid} are similar to our proposed method.
  %     It is our hornor to propose similar ideas on image compression at the same period with ~\cite{cheng2020learned, lee2019hybrid}.}
  % % \textcolor{red}{ Section XX describes ...}
  % \subsection{Inference Speed analysis}
  % In this section, we further analyse the decoding time of different methods based on deep learning.
  % The evaluation is conducted on GPU environmnet, RTX2080Ti.
  % In order to measure the overall improvement of our method, we apply the Bjontegaard delta bitrate(BDBR) metric.
  % BDBR refers to bitrate under equivalent PSNR.
  % We compute the relative bitrate improvement over JPEG~\cite{wallace1992jpeg}
  % As shown in Tab~\ref{tab:speed}, we benchmark the inference speed on $768\times512$ image from ~\cite{Kodak}.
  % It is easy to see that our method has comparable performance with ~\cite{minnen2018joint}, which is very slow.
  % At the same time, the inference speed of our method is close to the method of ~\cite{balle2018variational}, which is of vital importance to practical application.

  \begin{figure*}[htbp]
    \centering
    \includegraphics[width=1.0\linewidth]{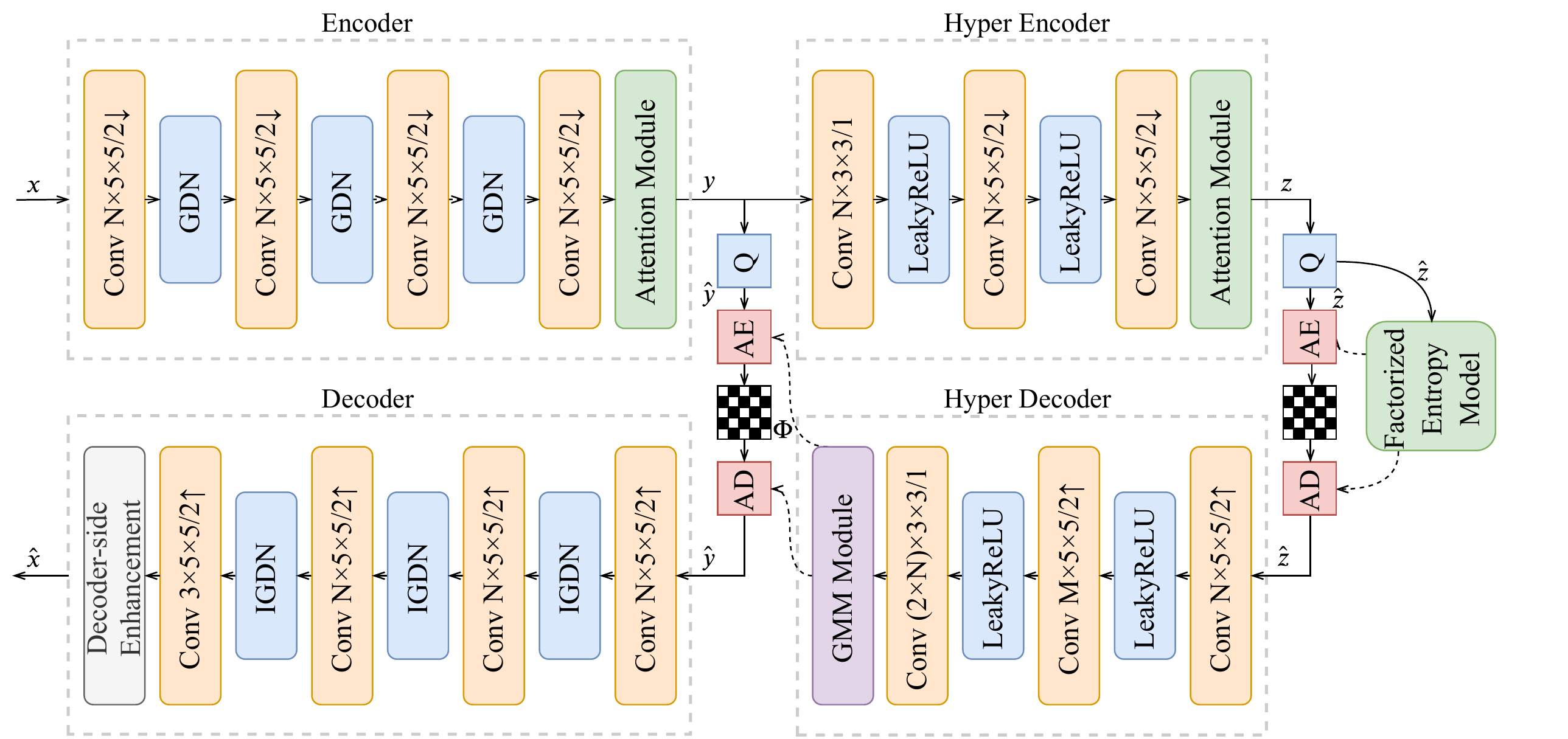}
    \caption{
       The framework of our proposed EDIC. Each convolution layer is denoted by the number of filters, kernel size, and stride.
       $\rightarrow$ indicates downsampling, and $\leftarrow$ indicates upsampling in each convolutional layer.
       $N$ and $M$ are the hyper-parameters to set the number of channels for a specific layer.
       ``GDN'' means generative divisive normalization proposed in~\cite{balle2015density}, and ``IGDN'' means inverse GDN.
       ``Q'' denotes quantization. ``AE'' and ``AD'' represent arithmetic encoder and arithmetic decoder, respectively.
       ``$\Phi$'' refers to the estimated parameters of the Gaussian mixture model.
       }
    \label{figure: architecture}
  \end{figure*}
      \section{Related Work}

      \subsection{Traditional Image and Video Compression}
      
      The image and video compression techniques are widely used to save the bandwidth and storage size in practical applications. 
      In the past decades, a lot of image and video compression methods have been proposed and several standards are also successfully built. 
      To improve the compression efficiency, the traditional image and video compression methods \cite{wallace1992jpeg,skodras2001jpeg,WebP,BPG} rely on manually designed techniques, such as liner transform and block based motion estimation and motion compensation schemes.  
      
      The image compression methods mainly focus on reducing the spatial redundancy in images. 
      One straightforward method is to convert the images from the pixel domain to the frequency domain, which is easier for compression. For example, the JPEG \cite{wallace1992jpeg} uses the discrete cosine transform while JPEG2000 \cite{skodras2001jpeg} employs discrete wavelet transform. 
      After the transform procedure, these coefficients are quantized, and then are sent to the decoder side. 
      To further improve the compression efficiency, the quantized coefficients are losslessly encoded by using the entropy coding tools, such as arithmetic coding~\cite{witten1987arithmetic}.
      Recently, the intra prediction technique in video compression is also exploited for image compression. For example, the BPG \cite{BPG} standard is based on HEVC/H.265 \cite{sullivan2012overview}, which achieves the state-of-the-art image compression performance when compared with the previous image codecs, such as JPEG and JPEG2000.
      The BPG standard adopts the prediction-transform technique and employs 35 encoding modes to obtain the predicted image, which further reduces the spatial redundancy. 
      
      Video compression is used to reduce the temporal redundancy in video sequences. Most video compression algorithms follow the hybrid coding architecture for high compression efficiency. In particular, H.264~\cite{x264} is the most widely used video codec. 
      In H.264, the block based motion estimation and motion compensation modules are utilized to obtain the predicted frame. Then we can calculate the residual information, which is compressed by using linear transform.
      Recently, HEVC/H.265 \cite{sullivan2012overview} and versatile video coding (VVC) are proposed as the next generation video codecs. 
      These standards build upon the previous hybrid coding architecture and utilize more advanced techniques for high efficiency coding.
      For example, HEVC uses the so-called Coding Unit (CU) Tree technique with the CU size ranging from $64 \times 64$  to $8 \times 8$, which provides flexible coding units for different video contents.

      \subsection{Learning based Image and Video Compression}
      
      In the past few years, deep neural network (DNN) has demonstrated its effectiveness for a lot of computer vision tasks, including super-resolution, denoising, etc. 
      Recently, researchers try to exploit the powerful representation ability from neural networks to enhance the image/video compression performance \cite{toderici2015variable,toderici2017full,balle2016end, balle2018variational, theis2017lossy, agustsson2017soft,li2017learning, rippel2017real,mentzer2018conditional,agustsson2018generative,minnen2018joint,lee2018context}.
      Toderici \textit{et al.} proposed the first learning based image compression framework by using recurrent neural network (RNN). Their approach can generate multiple bitrates through a single model. 
      In \cite{johnston2017improved}, more advanced RNN modules and effective reconstruction techniques are introduced to achieve comparable or even better performance when compared with BPG in terms of MS-SSIM~\cite{wang2003multi}
      However, these methods \cite{toderici2015variable,toderici2017full,johnston2017improved} are designed to minimize the bitrates instead of considering the rate-distortion trade-off.
      
      In \cite{balle2016end}, Ball{\'{e}} \textit{et al.} proposed a CNN based image compression framework by optimizing rate-distortion criterion.  To improve the accuracy of the entropy model, a hyper-prior model is proposed in \cite{balle2018variational}, where the latent representations are modeled based on zero-mean Gaussian distribution. 
      In \cite{minnen2018joint}, Minnen \textit{et al.} employed the auto-regressive priors to further improve the compression and achieve better performance than BPG in terms of PSNR. 
      However, these CNN based image compression systems have to train different models for different bitrates and increase the model sizes significantly. 
      In \cite{choi2019variable}, Choi \textit{et al.} proposed a variable rate deep image compression framework by using a conditional autoencoder and generates different bitrates through a single model.
      
      Considering that the quantization procedure itself is not differentiable, it is non-trivial to optimize the image compression system in an end-to-end manner.
      In \cite{balle2016end}, the quantization operation is approximated by adding uniform noise in the training stage. In \cite{theis2017lossy}, the gradients of quantization operation in the training stage are replaced for end-to-end optimization.
      To further improve the compression efficiency,  Rippel \textit{et al.} \cite{rippel2017real} used the multi-scale image decomposition technique to exploit the relationship between different scales. Agustsson \textit{et al.} \cite{agustsson2018generative} proposed a generative adversarial network based image compression system,
      which provides a visually pleasing reconstructed image for very low bitrate compression. 
      In addition, Li \textit{et al.} \cite{li2017learning} investigated the spatial relation in the latent representations and computed the importance map to guide the learning based image compression method.
      Inspired by the intra-prediction technique in traditional video coding, Baig \textit{et al.} used the inpainting method \cite{baig2017learning} to obtain the predicted block in the reconstructed frame and encode the corresponding residual by using neural network.  
      Although the current state-of-the-art learning based methods \cite{minnen2018joint,lee2018context} achieve better performance than the traditional methods such as BPG \cite{BPG}, the computational cost increases nearly 100 times when the auto-regressive prior \cite{minnen2018joint,lee2018context} is employed.
      Therefore, it is critical to build a more efficient image compression framework for practical applications.

      Recently, learning based video compression has attacted more and more attention.  Wu \textit{et al.} formulated video compression as frame interpolation and applied neural network to encode the residual information.
      Lu \textit{et al.} \cite{lu2019dvc} followed the traditional hybrid coding architecture and employed the neural networks to implement the video compression procedure, which can be optimized in an end-to-end manner.
      Cheng \textit{et al.} \cite{cheng2019learning} used the interpolation loop in the coding procedure and designed a spatial energy compaction-based penalty term into the loss function for better coding efficiency.
      In \cite{habibian2019video}, a 3D autoencoder scheme is proposed for video compression without computing the motion information. 
      In \cite{djelouah2019neural}, the proposed framework can decode the latent representations into motion and blending coefficients. Besides, the residual information is compressed in the latent space instead of the pixel domain.
  
  \section{Proposed Methods}
  \subsection{Overall Architecture for Image Compression}
  \label{sec:image}
  
  In this section, we introduce the proposed efficient deep image compression framework called EDIC. The architecture of the proposed scheme is illustrated in Fig.~\ref{figure: architecture}.
  Inspired by the recent progress in learning based image compression \cite{balle2016end,balle2018variational}, we also utilize the auto-encoder style network for learning based image compression.
  Specifically, there are four modules in the proposed scheme, \textit{i.e.,} encoder network, decoder network, hyper-encoder network, and hyper-decoder network.
  The encoder network takes the original image $x$ as the input and generates the corresponding latent representations $y$ by using several convolutional layers and non-linear functions. 
  The latent representations $y$ will be quantized to $\hat{y}$. Following arithmetic coding, like arithmetic encoder and arithmetic decoder, the quantized latent representations $\hat{y}$ were sent to the decoder network to reconstruct the final decoded image $\hat{x}$.
  We adopt the same quantization strategy as ~\cite{balle2018variational,minnen2018joint}.
  Considering that the image compression methods aim to achieve high quality reconstructed image at a given bitrate target and entropy model is used to estimate the bitrate,
  it is critical to build an accurate entropy model.
  In the proposed framework, we follow the pipeline in \cite{balle2018variational,minnen2018joint} and apply the hyper-encoder and hyper-decoder modules to estimate the parameters for the entropy model.
  Specifically, based on the latent representations $y$, the hyper-encoder module obtains the hyper-prior information and encodes it to latent representations $z$.
  Similarly, the latent representations $z$ will be quantized as $\hat{z}$.
  Then, quantized $\hat{z}$ will be sent by arithmetic coding.
  Finally, the hyper-decoder will reconstruct the hyper-prior information by using quantized hyper-latent representations $\hat{z}$ as the input and estimate the corresponding parameters $\Phi$ of the entropy model.
  The entropy model of hyper-latent representations is the same as ~\cite{balle2018variational,balle2016end}.
  The network architecture and entropy model in our proposed method will be discussed in the next three sections.
  
  The whole learning based image compression framework is optimized by considering the rate-distortion trade-off in the following way:
  
  \begin{equation}
      L = \lambda D +  R  = \lambda d(x, \hat{x}) + H(\hat{y}) + H(\hat{z}),
  \end{equation}
  where $D$ and $R$ represent the distortion and bitrate, respectively.
  $\lambda$ is the trade-off parameter.
  $d(\cdot)$ is the distortion metric (mean square error or MS-SSIM~\cite{wang2003multi}). 
  $H$ represents the bitrate for encoding latent representations $\hat{y}$ and $\hat{z}$.
  In the proposed method, the bitrate is approximated by using the entropy of the corresponding latent representations, \textit{i.e.,} $H(\hat{y}) = E[-log_2(p_{\hat{y}| \hat{z}}(\hat{y} | \hat{z}))]$ and $H(\hat{z}) = E[-log_2(p_{\hat{z}}(\hat{z}) )]$.
  $p_{\hat{y}| \hat{z}}(\hat{y} | \hat{z})$ and $p_{\hat{z}}(\hat{z}) $ represent the distributions of $\hat{y}$ and $\hat{z}$, respectively.

  \subsection{Channel Attention Scheme}
  \label{attention}
\begin{figure}[htbp]
    \centering
    \includegraphics[width=1.0\linewidth]{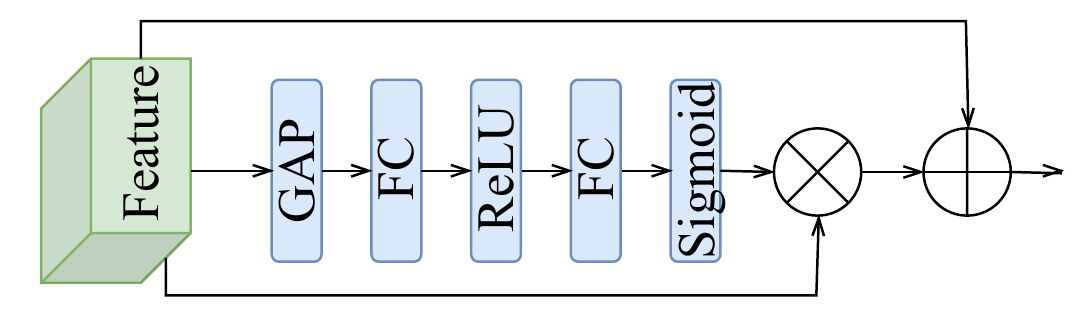}
    \caption{
        The structure of our channel attention module.``GAP'' represents global average pooling. ``FC'' means fully-connected layer.
       }
    \label{figure: attention}
  \end{figure}

  In \cite{minnen2018joint,lee2018context}, the auto-regressive prior model which captures the spatial relationship in latent representations is used to improve the compression performance. 
  Meanwhile, some works have applied spatial attention mechanisms implemented by non-local blocks~\cite{wang2018non} to image compression~\cite{chen2019neural, liu2019non}, which aims to reduce the spatial redundancy.
  Based on the aforementioned two motivations and inspired by~\cite{hu2018squeeze}, we propose to use a light-weight channel attention technique to exploit channel attention in the latent representations $\hat{y}$ and $\hat{z}$.
  The architecture of the proposed attention module is shown in Fig.~\ref{figure: attention}. 
  Let us denote the input feature map as $\mathbf{X}$, $\mathbf{X}\in\mathbb{R}^{{I}\times{J}\times{C}}$, where $I$, $J$, and $C$ denote height, width and channel dimension of the feature map, respectively.
  First, we apply global average pooling to obtain the channel-wise statistics $\mathbf{t}\in{\mathbb{R}^{C}}$, which is formulated bellow:
  \begin{equation}
    t_{c} = \frac{1}{{I}\times{J}}\sum_{i=1}^{I}\sum_{j=1}^{J}x_{c}(i, j),
  \end{equation}
  where $t_{c}$ means the ${c}$-th element of $\mathbf{t}$, and $x_{c}(i, j)$ represents the ${c}$-th channel specific value of the input feature map $\mathbf{X}$.
  Then, we apply several non-linear transforms to capture the channel-wise relationship. 
  Specifically, the non-linear transforms are described in the following formula:
  
  \begin{equation}
    \mathbf{s} = \sigma(\mathbf{W_{2}}\delta(\mathbf{W_{1}}\mathbf{t})),
  \end{equation}
  where $\mathbf{s}$ refers to the output channel-wise attention value, and $\mathbf{W_{1}}\in{\mathbb{R}^{\frac{C}{r}\times{C}}}$ and $\mathbf{W_{2}}\in{\mathbb{R}^{{C}\times\frac{C}{r}}}$ denotes the fully-connected layers,
  $\delta$ is the ReLU activation function~\cite{nair2010rectified} for non-linear transform, and $\sigma$ represents sigmoid activation.
  For reducing the dimension, we set $r$ as {16}. Finally, we re-scale the input feature map $\mathbf{X}$ with $\mathbf{s}$. In addition, we add the residual operation in our implementation.
  
  % The proposed module will take the features from the previous convolutional layer as input. Then the feature will go through the global average pooling, fully connected layer, Relu, etc, and we can compute the weights for each channel. Based on the estimated weights, the input feature maps are re-weighted for the following quantization and entropy coding. 
  
  As shown in Fig.~\ref{figure: architecture}, the proposed channel attention module is integrated into the encoder network and hyper encoder network and utilized to exploit the channel relationship for high quality compression. 
  We apply the re-weighted feature map to the following quantization and entropy coding modules.
\begin{figure}[htbp]
  \centering
  \includegraphics[width=0.32\linewidth]{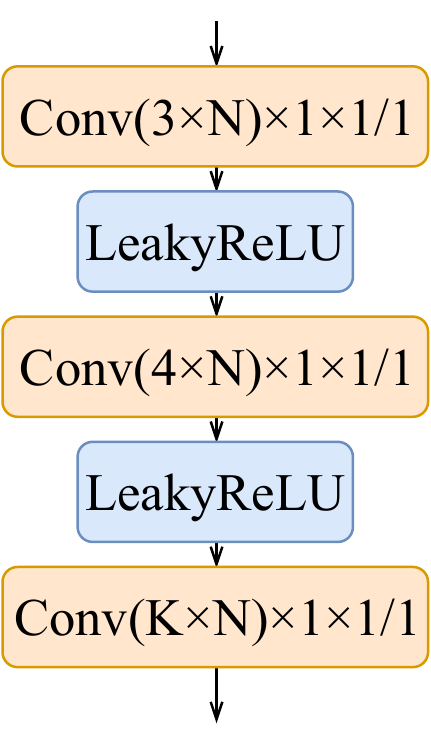}
  \caption{
    The structure of our GMM Module. $N$ denotes the hyper-parameter to set the number of channels for a specific layer,
    and $K$ depends on the number of Gaussian models. (See Section~\ref{sec:gmm} for more explanations)
     }
  \label{figure: gmm}
\end{figure}
\subsection{Gaussian Mixture Model for Entropy Estimation}
\label{sec:gmm}
\begin{figure*}[htbp]
  \centering
  % \subfigure[Kodim20.png]{
      % \centering

  \begin{minipage}[b]{1.0\textwidth}
      \includegraphics[width=0.24\textwidth]{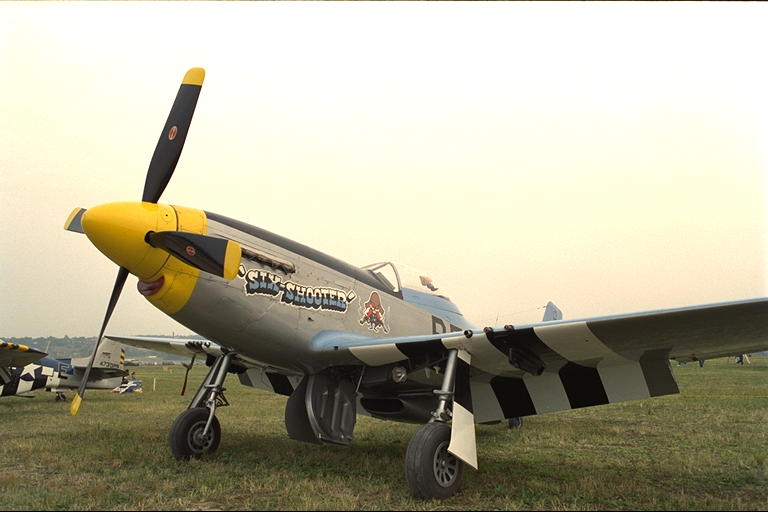} 
      \includegraphics[width=0.24\textwidth]{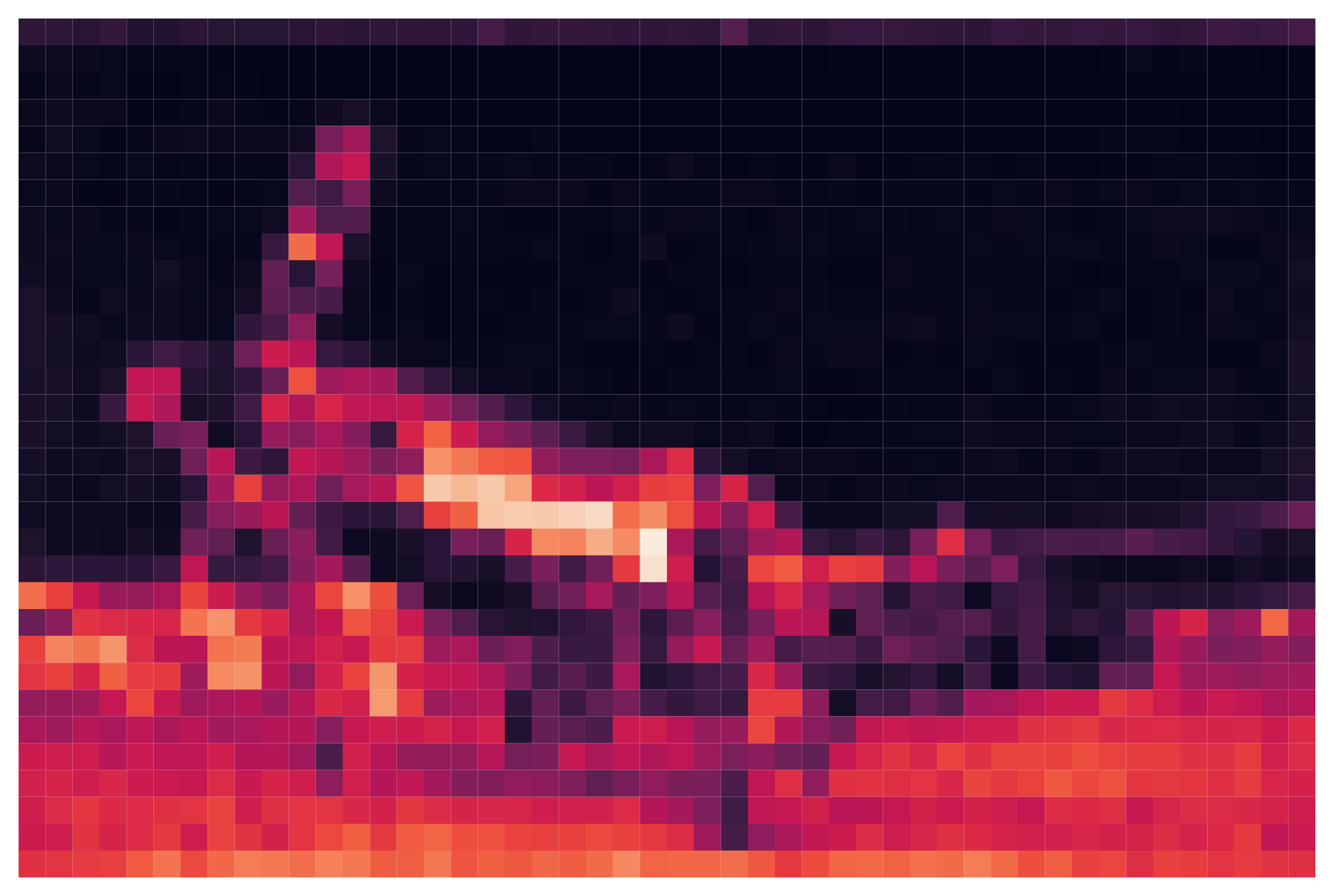} 
      \includegraphics[width=0.24\textwidth]{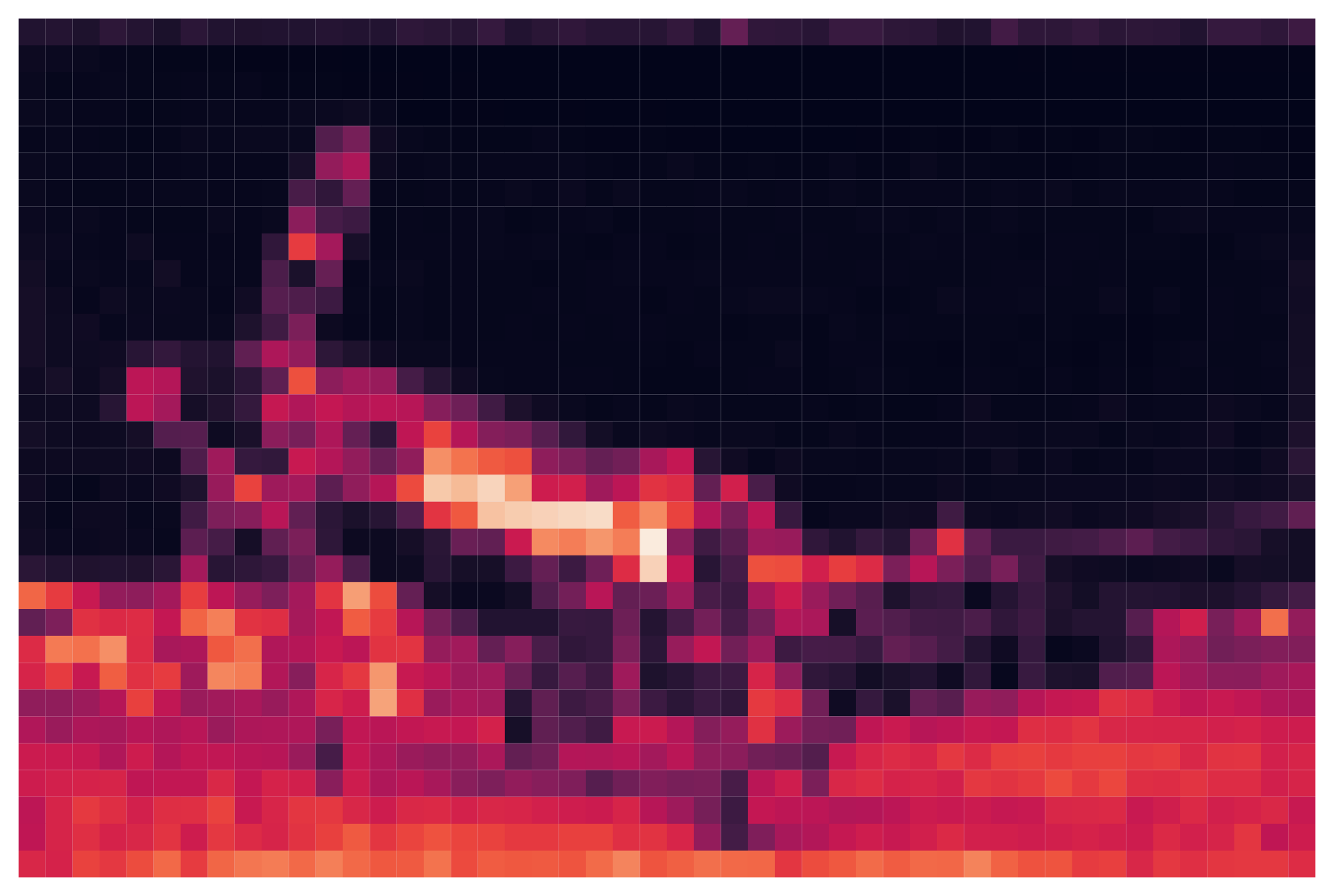}  
      \includegraphics[width=0.24\textwidth]{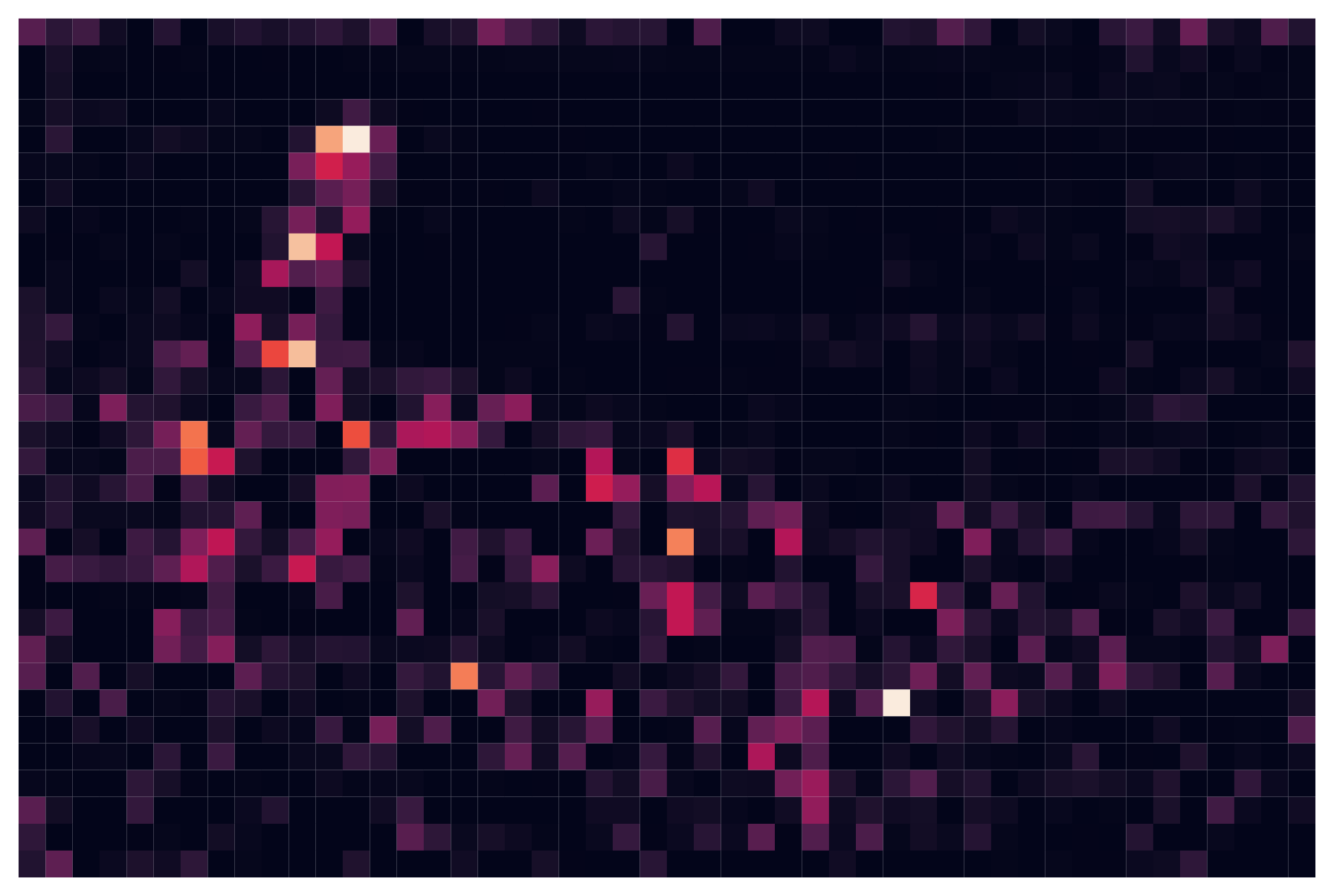}  
  \end{minipage}
  \\[2pt]
  % \subfigure[Kodim23.png]{
  %     \centering

  % \begin{minipage}[b]{1.0\textwidth}
  %     \includegraphics[width=0.24\textwidth]{images/kodim23.png}
  %     \includegraphics[width=0.24\textwidth]{images/gmm_vis_1/sg/23.pdf} 
  %     \includegraphics[width=0.24\textwidth]{images/gmm_vis_1/gmm/23.pdf}  
  %     \includegraphics[width=0.24\textwidth]{images/gmm_vis_1/diff/23.pdf}   
  % \end{minipage}
  \caption{Bit allocation map of the latent representations $\hat{y}$.
  The left column is the original image from Kodak~\cite{Kodak}. The middle left column is the bit allocation map of $\hat{y}$ after using a single Gaussian model as the entropy model.
  The middle right column is the bit allocation map of $\hat{y}$ after using the Gaussian mixture model as the entropy model.
  The right column is bit allocation difference between them. We take ``kodim20.png'' for visualization.}
  \label{fig:gmm}
\end{figure*}
In the learning based image compression methods, accurate bit rate estimation is critical. In \cite{minnen2018joint,lee2018context}, the learning based systems adopt the hyper-prior compression scheme and the latent representations $\hat{y}$ are modeled as Gaussian distribution as follows:

\begin{equation}
    p_{\hat{y}| \hat{z}}(\hat{y} | \hat{z}) \sim \mathcal{N}(\mu, \sigma),
\end{equation}
where $p_{\hat{z}}(\hat{z})$ is represented by using the factorized entropy model~\cite{balle2016end}. 
The goal of hyper-encoder and hyper-decoder is to estimate the parameters  $\mu$ and $\sigma$ of the Gaussian model. 

Although the single Gaussian based entropy model has achieved significant improvements when compared with the previous work~\cite{balle2016end},
the representation ability of a single Gaussian model is still limited, especially for the complex contents.
Therefore, we utilize the Gaussian mixture model to further improve the efficiency of the image compression system. 
Specifically, the distribution of $\hat{y}$ is formulated as follows:
\begin{equation}
    p_{\hat{y}| \hat{z}}(\hat{y} | \hat{z}) \sim \sum_{i=1}^{F} \omega_i \mathcal{N}(\mu_i, \sigma_i),
\end{equation}
where $\omega_i$ represents the weights for different Gaussian models. $F$ is the number of Gaussian models.
As shown in Fig.~\ref{figure: gmm}, we design three convolutional layers with two LeakyReLU layers to estimate the parameters $\Phi$ of the Gaussian mixture model.
In our implementation, $F$ is set as 2. 
So the output channel number $K$ of the GMM module is set as $5\times{N}$,
the first $4\times{N}$ channels are used to estimate the mean and variance of two Gaussian models, respectively. In order to estimate the weights of each gaussian model. We add a sigmoid layer on the output of the last $N$ channels.
If the weight of one Gaussian model is $w$, the weight of another Gaussian model is $(1-w)$. 
Specifically, if we design ${F}(F\geq 3)$ Gaussian models, we can change the number of output channels of the GMM module to $3\times{F}\times{N}(K=3\times{F})$.
Similarly, the first $2\times{F}\times{N}$ channels estimate the mean and variance parameters of $F$ Gaussian models.
In particular, we add the softmax layer after the last $F\times{N}$ channels to calculate the weight of each Gaussian model. 

We provide further analysis of the GMM module. 
As shown in Fig.~\ref{fig:gmm}, the left part is the original image from Kodak~\cite{Kodak},
 the right part shows the estimated bit allocation map difference of the latent representations $\hat{y}$ between the single Gaussian model and Gaussian mixture model.
The brighter region indicates that the Gaussian mixture model saves more bits. 
From Fig.~\ref{fig:gmm}, it is clear that the Gaussian mixture model can save more bits, especially in the edge regions.
 
%  \begin{figure*}[htbp]
%   \centering
%   \subfigure[Kodim20.png]{
%       \centering

%   \begin{minipage}[b]{1.0\textwidth}
%       \includegraphics[width=0.24\textwidth]{images/kodim20.png} 
%       \includegraphics[width=0.24\textwidth]{images/gmm_vis/sg/20.pdf} 
%       \includegraphics[width=0.24\textwidth]{images/gmm_vis/gmm/20.pdf}  
%       \includegraphics[width=0.24\textwidth]{images/gmm_vis/diff/20.pdf}  
%   \end{minipage}}
%   \subfigure[Kodim23.png]{
%       \centering

%   \begin{minipage}[b]{0.5\textwidth}
%       \includegraphics[width=0.49\textwidth]{images/kodim23.png}
%       \includegraphics[width=0.49\textwidth]{images/1.pdf}  
%   \end{minipage}}
%   \caption{Effiectiveness of Gaussian mixture model.
%   For visualization, we adopt ``Kodim20.png'' and ``Kodim23.png'' from ~\cite{Kodak}.}
%   \label{fig:gmm}
% \end{figure*}

\subsection{Decoder-side Enhancement}
\label{decoder}
\begin{figure}[htbp]
  \centering
  \includegraphics[width=0.7\linewidth]{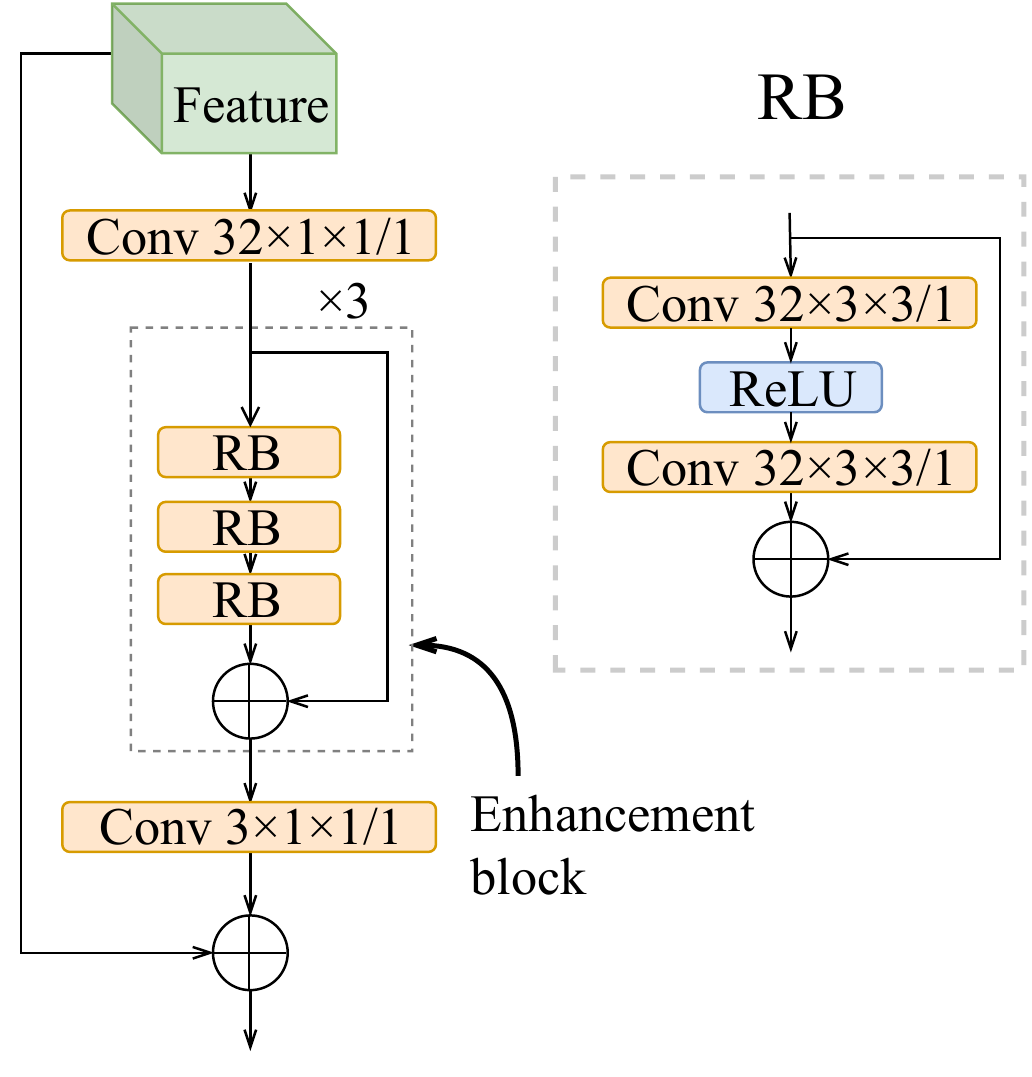}
  \caption{
    The structure of our decoder-side enhancement module. ``RB'' refers to the residual block.
     }
  \label{figure: enhance}
\end{figure}

\begin{figure}
  \centering
  % \subfigure[Bit allocation map.]{
      % \centering
  \begin{minipage}[b]{0.45\textwidth}
    \centering
    \includegraphics[width=0.48\textwidth]{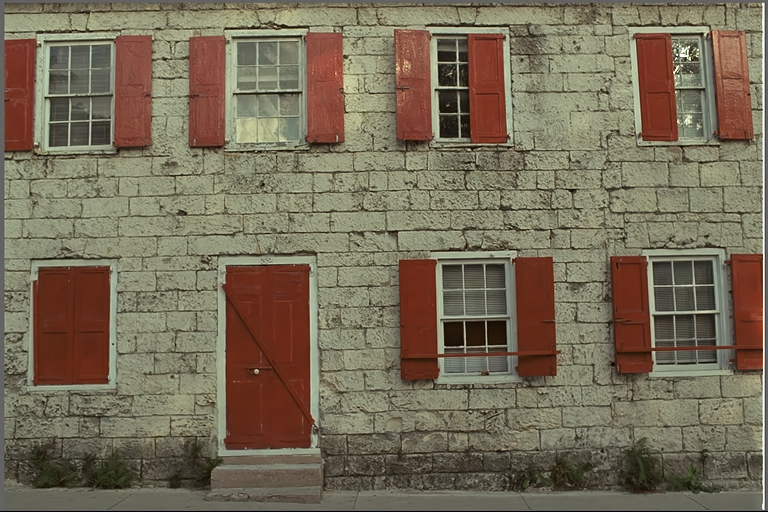}
    \includegraphics[width=0.48\textwidth]{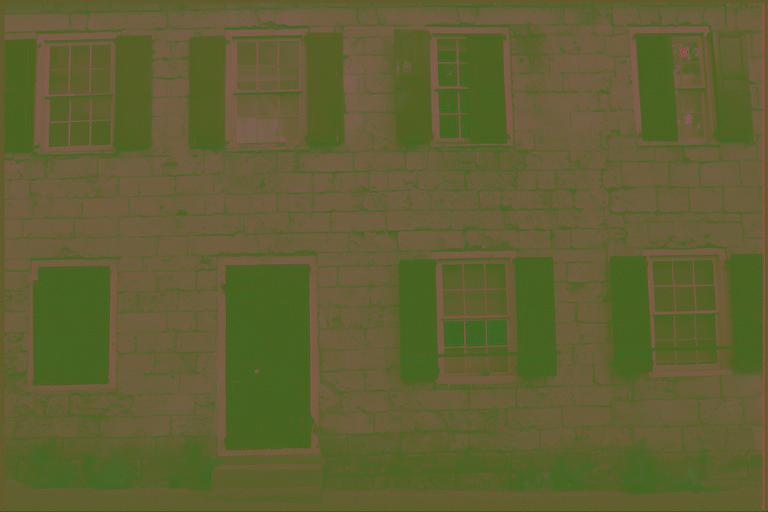}
  \end{minipage}
  % }
  \caption{The left image is the reconstructed image after the decoder-side enhancement module,
  and the right image the learned residual image.
  We take ``kodim01.png'' from Kodak~\cite{Kodak} for illustration.}
  \label{fig:enhance_vis}
\end{figure}

Since the proposed compression scheme is a lossy procedure, the reconstructed image has compression artifacts inevitably. 
To further improve the reconstructed quality, we introduce an enhancement module at the decoder side after image reconstruction. 
We adopt several residual blocks to restore the original image based on the input reconstructed image.
Inspired by the network design strategy for super resolution~\cite{lim2017enhanced},
we introduce the residual block to learn the high frequency information for image compression.
As shown in Fig.~\ref{figure: enhance}, we first add a convolution layer to increase the channel dimension from 3 to 32.
Then, we apply three enhancement blocks to the output of the convolution layer.
Every enhancement block has three residual blocks.
Finally, we add a convolution layer and residual operation to obtain the reconstructed image.
Moreover, the decoder-side enhancement module can be readily integrated into the whole compression system and optimized in an end-to-end manner with high efficiency.
As shown in Fig.~\ref{fig:enhance_vis},
we provide analysis about the decoder-side enhancement module.
The learned image is the output after the final convolution layer.
We observe that the learned residual image mainly contains the high frequency information,
which means that the decoder-side enhancement module helps to predict the high frequency components.

\subsection{Extension for Video Compression}

In order to further demonstrate the effectiveness of our newly proposed method,
we also apply our proposed method for the video compression task.
In our work, we choose DVC~\cite{lu2019dvc} as our baseline algorithm.

\begin{figure}[htbp]
  \centering
  \includegraphics[width=1.0\linewidth]{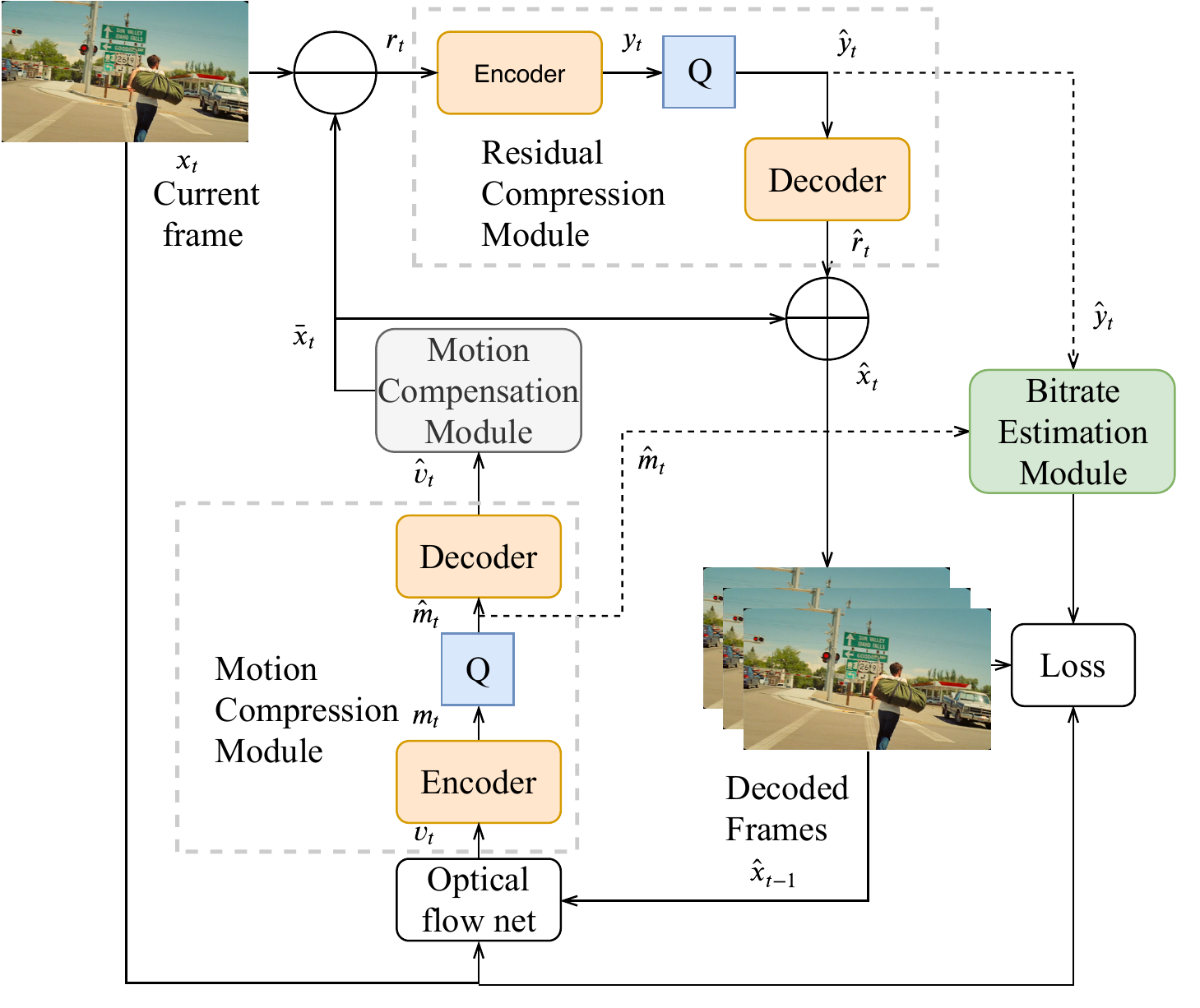}
  \caption{
    The framework of our video compression method.
    The network structures of the residual compression module and the motion compression module are the same as in the Fig.~\ref{figure: architecture}.
    The bitrate estimation module is our method for estimating the bitrate of the latent representations.
    ``Q'' denotes quantization.
     }
  \label{figure: video}
\end{figure}

The overall framework is shown in Fig.~\ref{figure: video}.
\{$x_{1}, x_{2},...,x_{t-1}, x_{t}$\} denote the current video sequences.
$x_{t}$ refers to the frame at time-step $t$.
$\hat{x}_{t}$ represents the reconstructed frame.
$m_t$ and $r_t$ are the motion and residual information, respectively.
The procedure of our video compression framework is shown as follows.
\subsubsection{Motion Estimation and Compression}
We utilize the CNN model proposed by~\cite{Ranjan_2017_CVPR} to predict optical flow,
which represents the motion information $v_{t}$.
Instead of encoding the motion information $v_{t}$ directly,
we send $v_{t}$ to the encoder network of the motion compression module to obtain $m_{t}$,
Then, we will quantize $m_{t}$ and reconstruct the motion information $\hat{v}_{t}$ by using the decoder network of the motion compression module.
\subsubsection{Motion Compensation, Residual Compression and Frame Reconstruction}
The motion compensation module takes the previous reconstructed frame $\hat{x}_t$ and motion information $\hat{v}_{t}$ as the input,
and obtains the predicted frame $\bar{x}_{t}$, which is supposed to be as close to the current frame $x_{t}$ as possible.
After that, we use the original frame $x_{t}$ and $\bar{x}_{t}$ to obtain residual information $r_{t}$, where $r_{t}=x_{t}-\bar{x}_{t}$.
The encoder network of the residual information module encodes the resiudal information $r_{t}$,
and quantizes $r_{t}$ to obtain the latent representations $y_{t}$.
Similarly, the decoder network of the residual information module reconstructs the residual information $\hat{r}_{t}$.
Then, the final reconstructed frame $\hat{x}_{t}$ can be obtained, where $\hat{x}_{t}=\hat{r}_{t}+\bar{x}_{t}$.

\subsubsection{Optimization of the framework}
The overall framework is optimized by minimizing the following Rate-Distortion trade-off:
\begin{equation}
  L_{t} = \lambda D_{t} +  R_{t}  = \lambda d(x_{t}, \hat{x}_{t}) + H(\hat{r}_{t}) + H(\hat{m}_{t}),
\end{equation}
where $L_{t}$ is the loss at the current time step $t$,
$d(\cdot,\cdot)$ is the distortion between the current frame $x_{t}$ and the reconstructed frame $\hat{x}_{t}$,
and $H(\hat{r}_{t})$ and $H(\hat{m}_{t})$ are the bitrates of the latent representations $\hat{r}_{t}$ of residual information and the latent representations $\hat{m}_{t}$ of motion information,
which are estimated by the bitrate estimation module.
% It is noticed that the whole network is optimized to minimize the rate-distortion criterion for the current time step,
% while the potential influence from to the next frame is ignored.

DVC utilizes the method proposed by Ball{\'{e}}~\cite{balle2018variational} to compress the residual information, 
and Ball{\'{e}}'s method~\cite{balle2016end} to compress the motion information.
In our work,
we propose to use our proposed EDIC image compression framework to compress both the residual and motion information.
Specifically, 
in the encoder network of the residual compression module and the motion compression module,
we utilize the proposed channel attention scheme described in Section~\ref{attention} to reduce the redundancy of the latent representations of residual and motion information.
In terms of bitrate estimation module,
we introduce the newly proposed Gaussian mixture model as the entropy model described in Section~\ref{sec:gmm} to estimate the bitrates of the latent representations more accurately,
in which the hyper-encoder and hyper-decoder network are used to estimate the parameters of Gaussian mixture model.
Furthermore, in the decoder network of the residual compression module and the motion compression module,
we add the decoder-side enhancement module in Section~\ref{decoder} to improve the reconstructed qualities of the residual and motion information effectively.

\section{Experiments}
In this section, we perform extensive experiments to demonstrate the effectiveness of our proposed EDIC framework,
which consists of the attention module, the GMM module and the decoder-side enhancement module.
With regard to image compression, we adopt $20745$ high-quality images {from Flick.com}\footnote{We have released our training data at~\url{https://github.com/liujiaheng/CompressionData}.} and randomly take ${256\times256}$ cropped patches for training.
For performance evaluation, we calculate the Rate-Distortion (RD) performance, which is averaged over all images in the Kodak PhotoCD image dataset~\cite{Kodak}.
For video compression, 
we use Vimeo-90k~\cite{xue2017video} dataset, which has 89,800 video clips with the resolution of ${256}\times{256}$, as our training dataset,
and evaluate our model on the HEVC Standard Test Sequences (\textit{i.e.}, Class B, Class C, Class D,
Class E)~\cite{sullivan2012overview}, which is widely used for evaluating video compression methods.
Our EDIC framework is implemented on the PyTorch~\cite{NIPS2019_9015} platform.
All the experiments are conducted on the GPU NVIDIA 2080Ti server with 11 GB memory.
\subsection{Performance and implementation details for Image Compression}
\label{details}
% Details of the neural network are outlined in Table~\ref{tab:booktabs}.
For image compression with the quality metric as the MSE loss function,
we train our model using different $\lambda$ values (\textit{i.e.}, 256, 512, 1024, 2048, 4096, 6144 8192).
In the first stage, we train the high bitrate point in the Rate-Distortion (RD) curve with $\lambda$ as 8192.
The model is trained on 1 GPU with the batch size of 4.
We apply Adam optimizer~\cite{kingma2014adam} with the learning rate of ${1\times10^{-4}}$ in the first 3,000,000 iterations and ${1\times10^{-5}}$ in the remaining 500,000 iterations.
For other bitrates, we just adopt the model trained on high bitrate ($\lambda=8192$) as a pre-trained model
and fine-tune our model.
We use Adam optimizer with the learning rate of ${1\times10^{-4}}$ in the first 500,000 iterations, and ${1\times10^{-5}}$ in the remaining 500,000 iterations.
Other training settings remain the same.
When our model is optimized with other quality metrics, such as the MS-SSIM loss function, we adopt the model optimized by the MSE loss function with $\lambda$ of 8192 as our pre-trained model.
Then, we change the MSE loss function to the MS-SSIM loss function and fine-tune the pre-trained model with different $\lambda$ values (\textit{i.e.}, 16, 32, 64, 128, 256, 384, 512).
We train the model with the learning rate of ${1\times10^{-5}}$ for 500,000 iterations.
Besides, we set $N$ to $320$ and $M$ to $480$.
% \begin{table*}
%     \centering
%     \begin{tabular}{lllll}  
%     \toprule
%     Encoder  & Decoder & Hyper Encoder & Hyper Decoder & GMM Module\\
%     \midrule
%     Conv: 5 $\times$ 5 c320 s2       & Deconv: 5 $\times$ 5 c320 s2  & Conv: 3 $\times$ 3 c320 s1  &Deconv: 5 $\times$ 5 c320 s2  &  Conv: 1 $\times$ 1 c960 s1\\
%     GDN           & IGDN  & LeakyReLU  & LeakyReLU  & LeakyReLU   \\ 
%     Conv: 5 $\times$ 5 c320 s2           & Deconv: 5 $\times$ 5 c320 s2  &Conv: 5 $\times$ 5 c320 s2 &Deconv: 3 $\times$ 3 c480 s1     &Conv: 1 $\times$ 1 c1280 s1 \\
%     GDN   & IGDN & LeakyReLU    & LeakyReLU  & LeakyReLU\\
%     Conv: 5 $\times$ 5 c320 s2           & Deconv: 5 $\times$ 5 c320 s2  &Conv: 5 $\times$ 5 c320 s2 &Deconv: 3 $\times$ 3 c640 s1     &Conv: 1 $\times$ 1 c1600 s1 \\
%     GDN   & IGDN     \\   
%     Conv: 5 $\times$ 5 c320 s2    & Deconv: 5 $\times$ 5 c3 s2     \\
%     \bottomrule
%     \end{tabular}
%     \caption{Network details of our framework. Convlution layer is denoted as: kernel size, number of filters and down- or upsampling stribe. GDN means generalized divisive normalization,
%     and IGDN means inverse GDN~\cite{balle2015density}.}
%     \label{tab:booktabs}
%     \end{table*}

\begin{figure}[htp]
  \centering
  \includegraphics[width=1.0\linewidth]{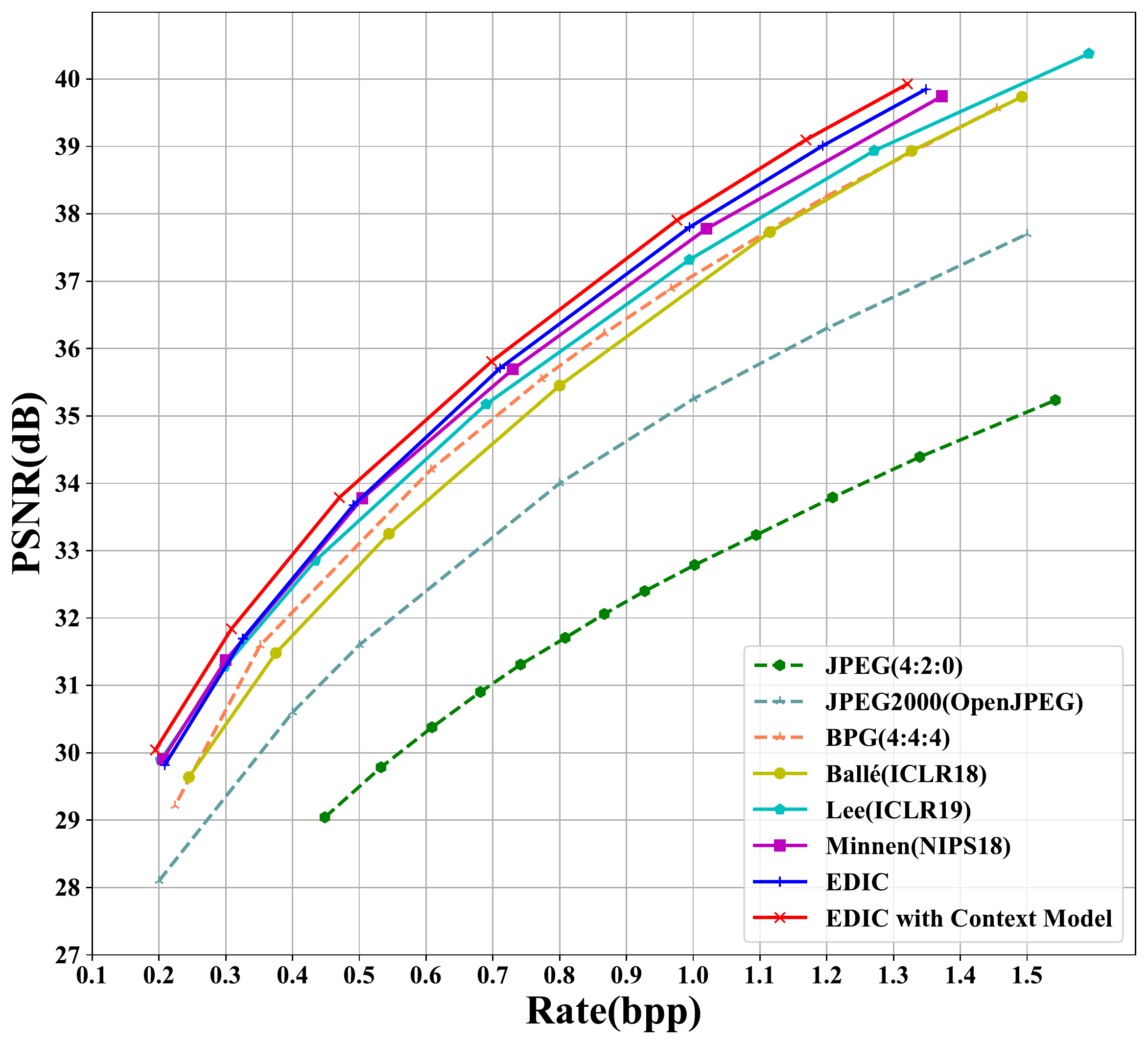}
  \caption{
    Rate-distortion curves of our proposed EDIC method and the competitive methods for image compression when using the PSNR metric. The ``Context Model'' is from Minnen's work~\cite{minnen2018joint}, which must be executed sequentially in the inference stage.
  }
  \label{figure: result}
\end{figure}

\begin{figure}[htp]
  \centering
  \includegraphics[width=1.0\linewidth]{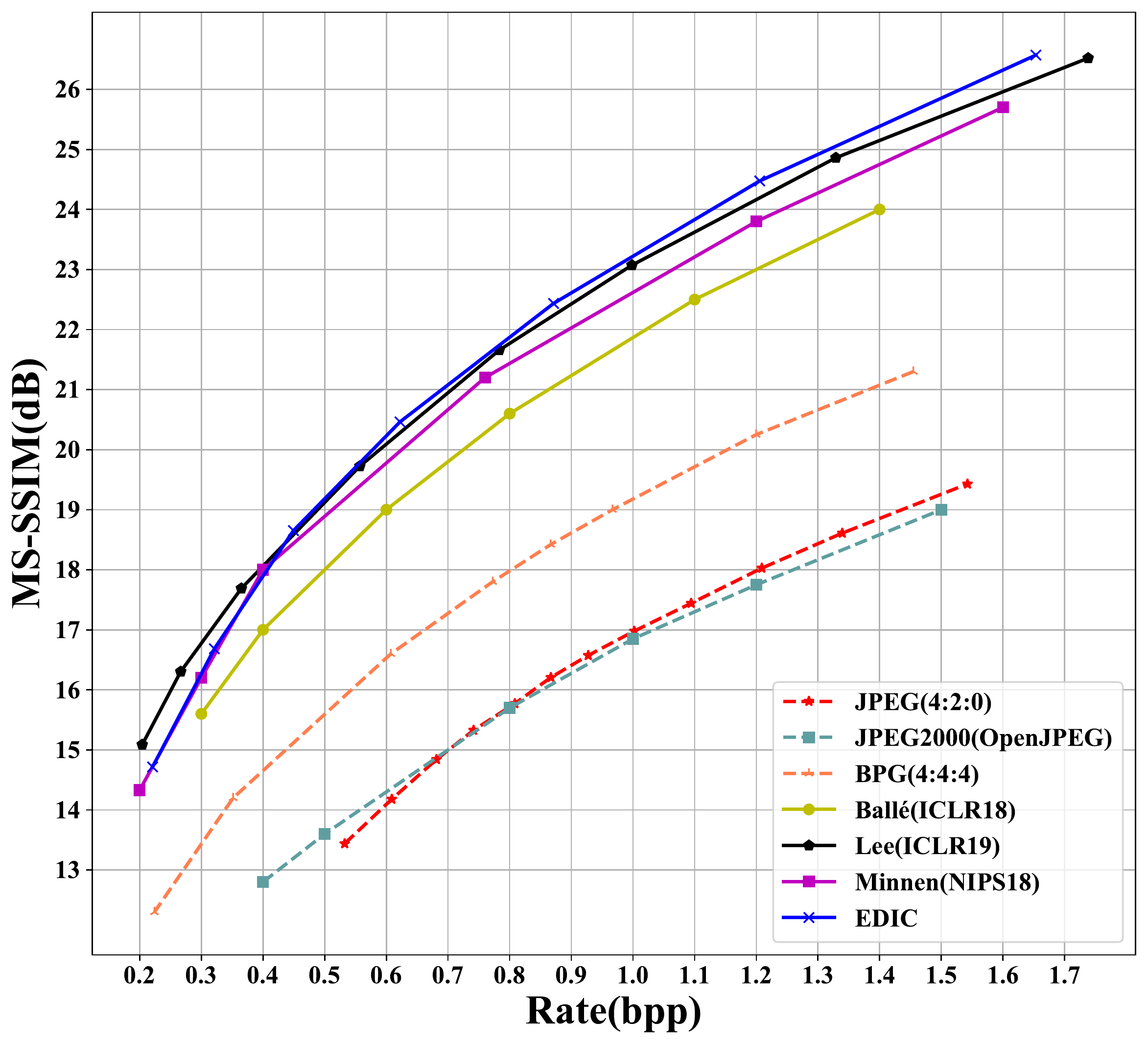}
  \caption{
    Rate-distortion curves of our proposed EDIC method and the competitive methods for image compression when using the MS-SSIM metric.
  }
  \label{figure: result-ssim}
\end{figure}

As shown in Fig.~\ref{figure: result}, we adopt peak signal-to-noise ratio (PSNR) as the quality metric.
We compare our EDIC method with the well-kown image compression standards, like BPG~\cite{BPG}, JPEG~\cite{wallace1992jpeg}, JPEG2000~\cite{skodras2001jpeg}, and recent neural networks methods,
like Ball{\'{e}}'s work~\cite{balle2018variational}, Minnen's work~\cite{minnen2018joint} and Lee's work~\cite{lee2018context}.
The results of Lee's work~\cite{lee2018context} are from their released source code$\footnote{https://github.com/JooyoungLeeETRI/CA\_Entropy\_Model}$.
The results of Ball{\'{e}}'s work~\cite{balle2018variational} and Minnen's work~\cite{balle2018variational} are based on our implementation.
When compared with the traditional methods, our EDIC has surpassed BPG~\cite{BPG}, JPEG~\cite{wallace1992jpeg}, JPEG2000~\cite{skodras2001jpeg} by a large margin.
When compared with the existing deep learning based methods, 
our EDIC achieves significant improvement over Ball{\'{e}}'s work~\cite{balle2018variational}.
As far as we know, the method proposed by Minnen~\textit{et al.} has achieved the state-of-the-art performance for image compression.
Our method has comparable results with Minnen's work~\cite{minnen2018joint} and Lee's method~\cite{lee2018context} at low bitrates,
and achieves apparent performance improvement over Minnen's work~\cite{minnen2018joint} and Lee's method~\cite{lee2018context} at high bitrates.
In addition, Minnen's work and Lee's method are very slow, because their inference strategies are sequential.
By contrast, our method can be readily parallelized.
As a result, our method is very efficient, which is very important for practical application scenarios.
Furthermore, the attention module, the GMM module, and the Decoder-side Enhancement module are all independent modules and can be easily incorporated with other methods.
As shown in Fig.~\ref{figure: result}, when we incorporate our method into Minnen's work~\cite{minnen2018joint},
which has the context model for estimating more accurate entropy parameters,
our EDIC method with context model also achieves over 0.2 dB improvement when compared with our EDIC method. 
which again demonstrated the effectiveness of our proposed schemes.
As shown in Fig.~\ref{figure: result-ssim},
we also conduct the experiments in terms of the MS-SSIM quality metric.
In order to describe the improvement more clearly, 
we report the MS-SSIM values using decibels $(\textit{i.e.}, -{10}\log_{10}(1-${MS-SSIM}$))$.
It is clear that our EDIC is better than BPG~\cite{BPG}, JPEG~\cite{wallace1992jpeg}, JPEG2000~\cite{skodras2001jpeg}, and Ball{\'{e}}'s~\cite{balle2018variational}.
When compared with the state-of-the-art methods, our EDIC is comparable with Minnen's method~\cite{minnen2018joint} and lower than Lee's work~\cite{lee2018context} at low bitrates.
However, our EDIC is apparently superior to their methods at high bitrates.
\begin{figure*}[htp]
  \centering
  % \subfigure[Rate-distortion curves of our method and competetive methods based on PSNR quality metric.]{
  %     \centering

  \begin{minipage}[b]{0.85\textwidth}
      \includegraphics[width=0.5\textwidth]{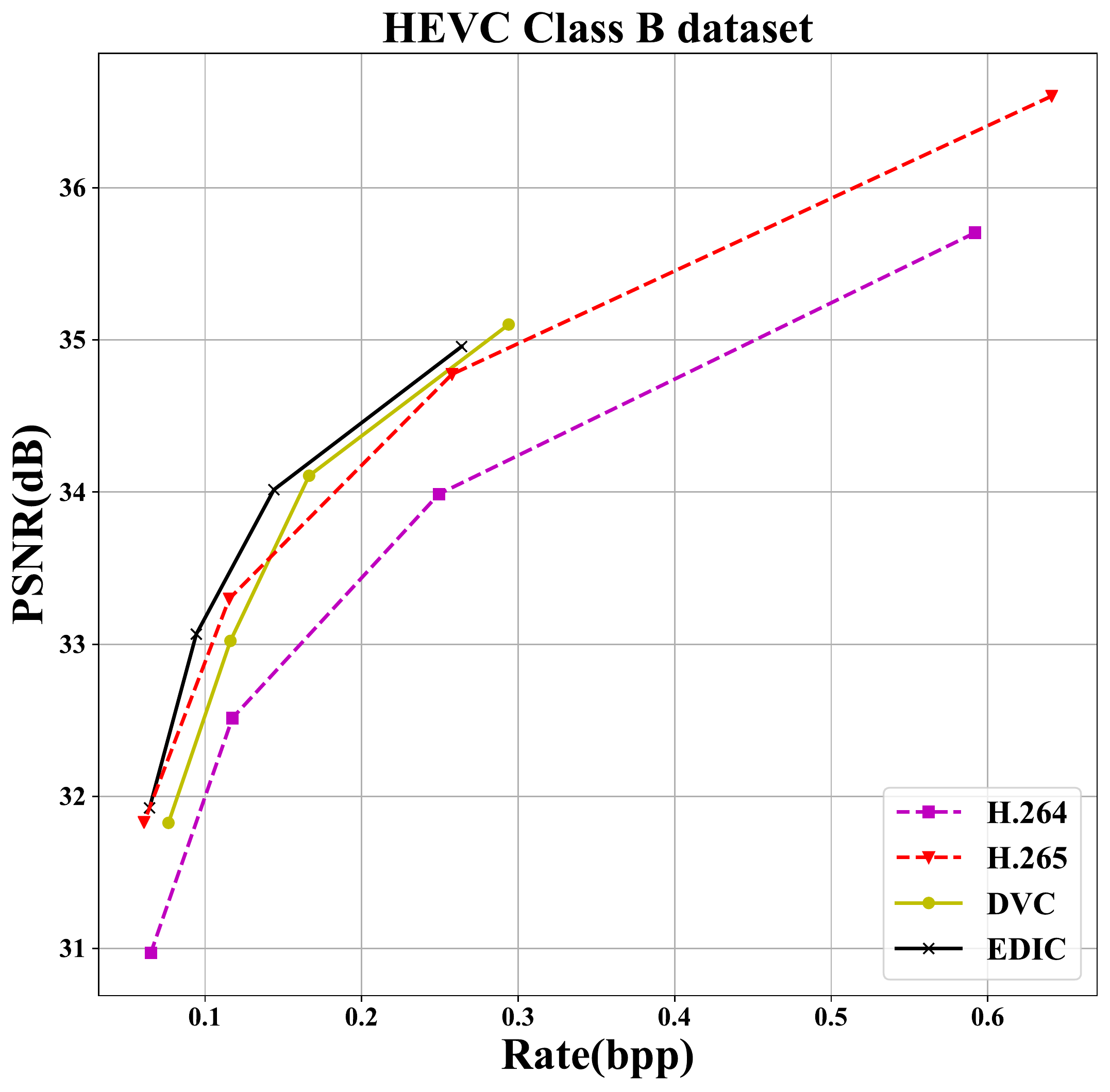} 
      \includegraphics[width=0.5\textwidth]{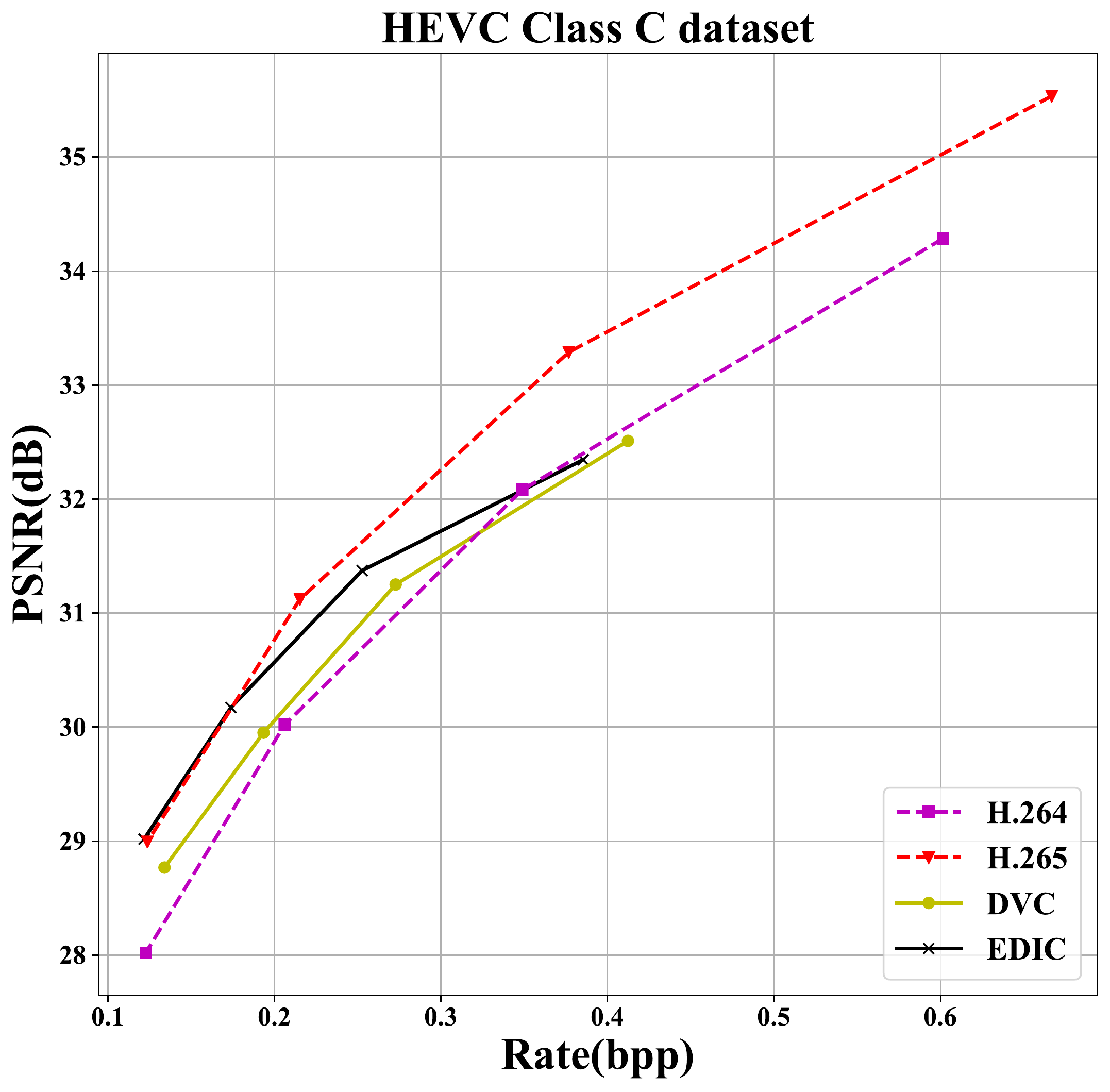} 
      \includegraphics[width=0.5\textwidth]{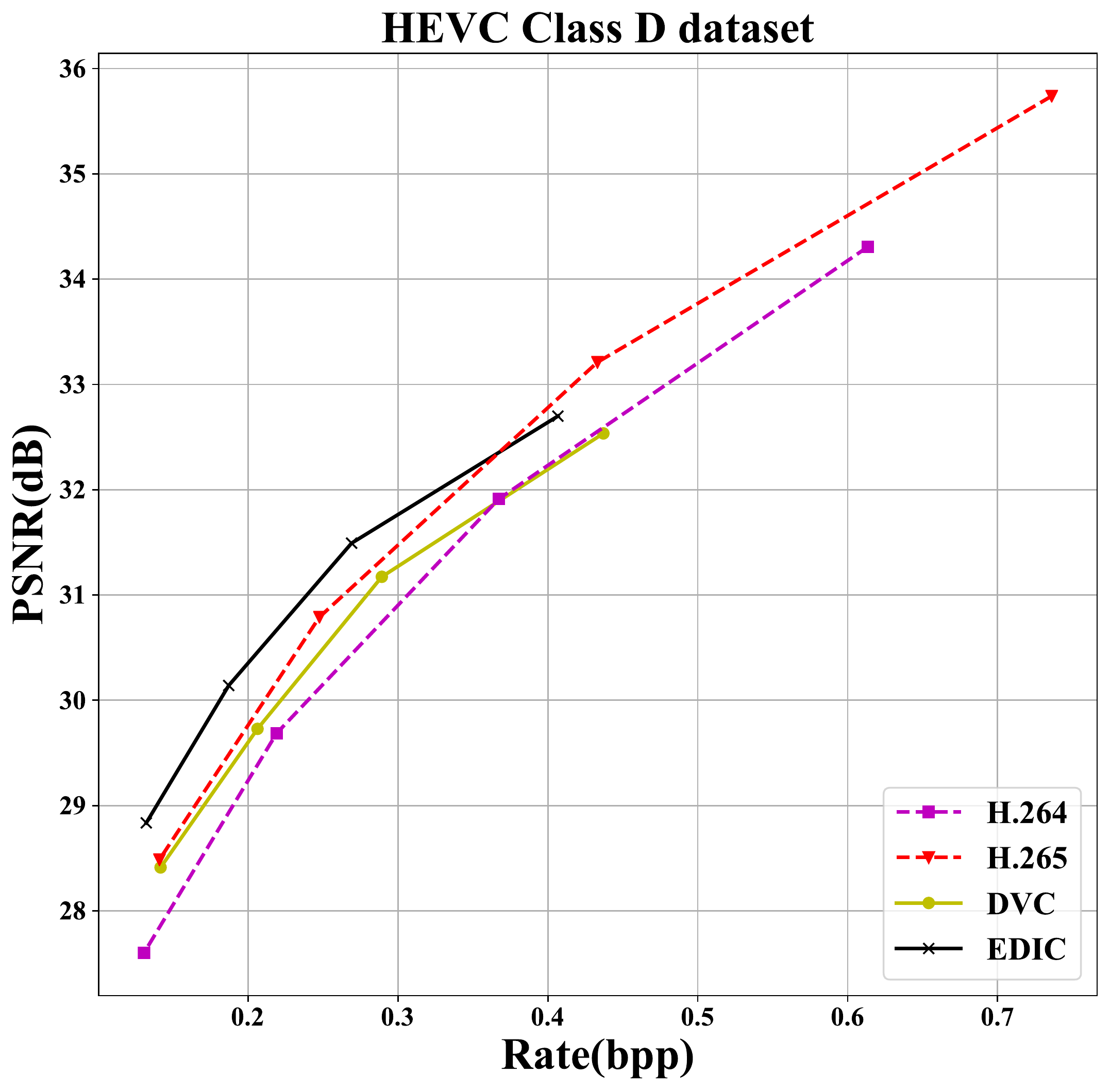}
      \includegraphics[width=0.5\textwidth]{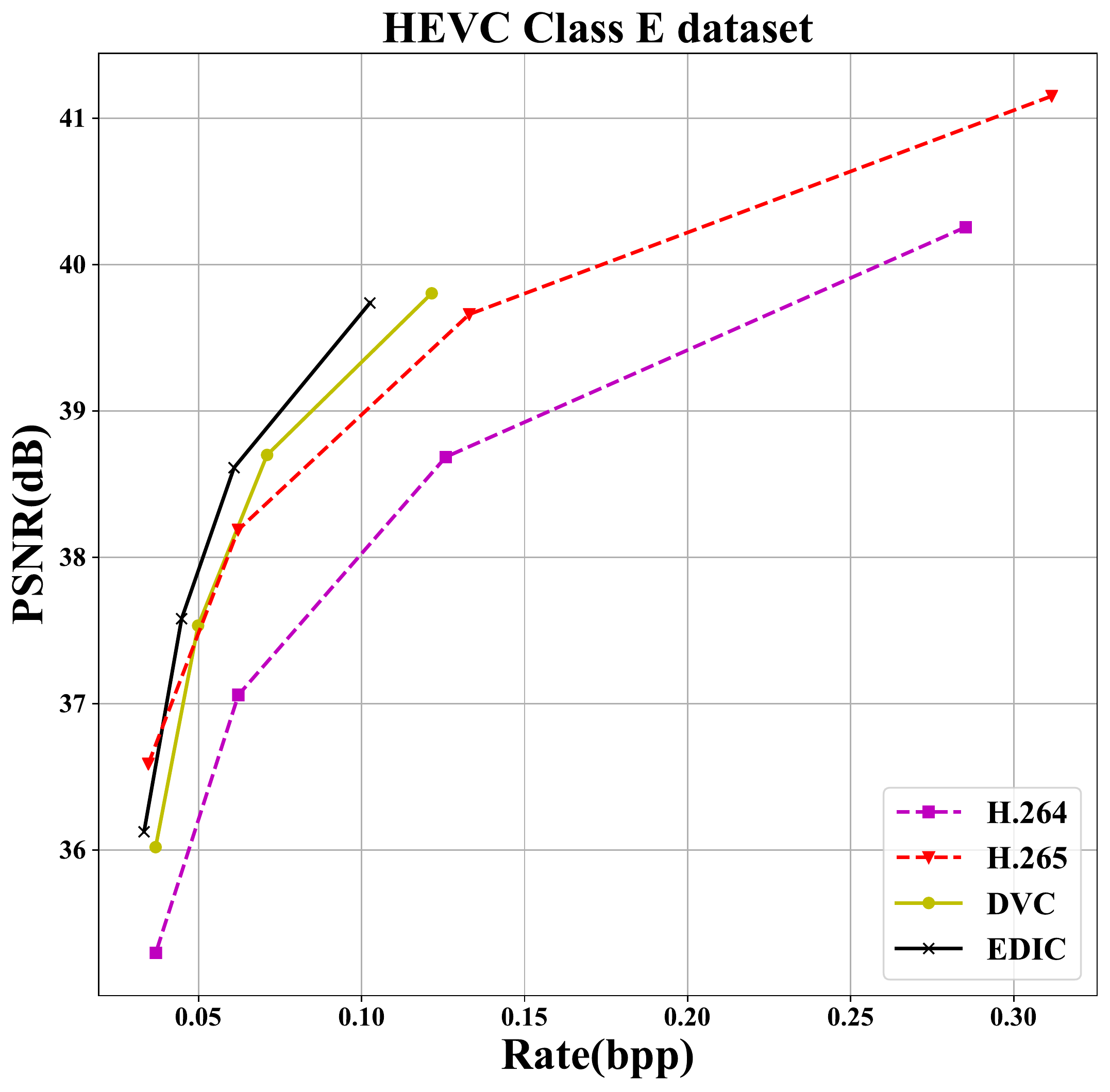}    
  \end{minipage}
  \caption{Rate-distortion curves of our proposed EDIC method and the competitive methods for video compression when using the PSNR metric.}
\end{figure*}
\begin{figure}[htp]
  \centering
  \includegraphics[width=1.0\linewidth]{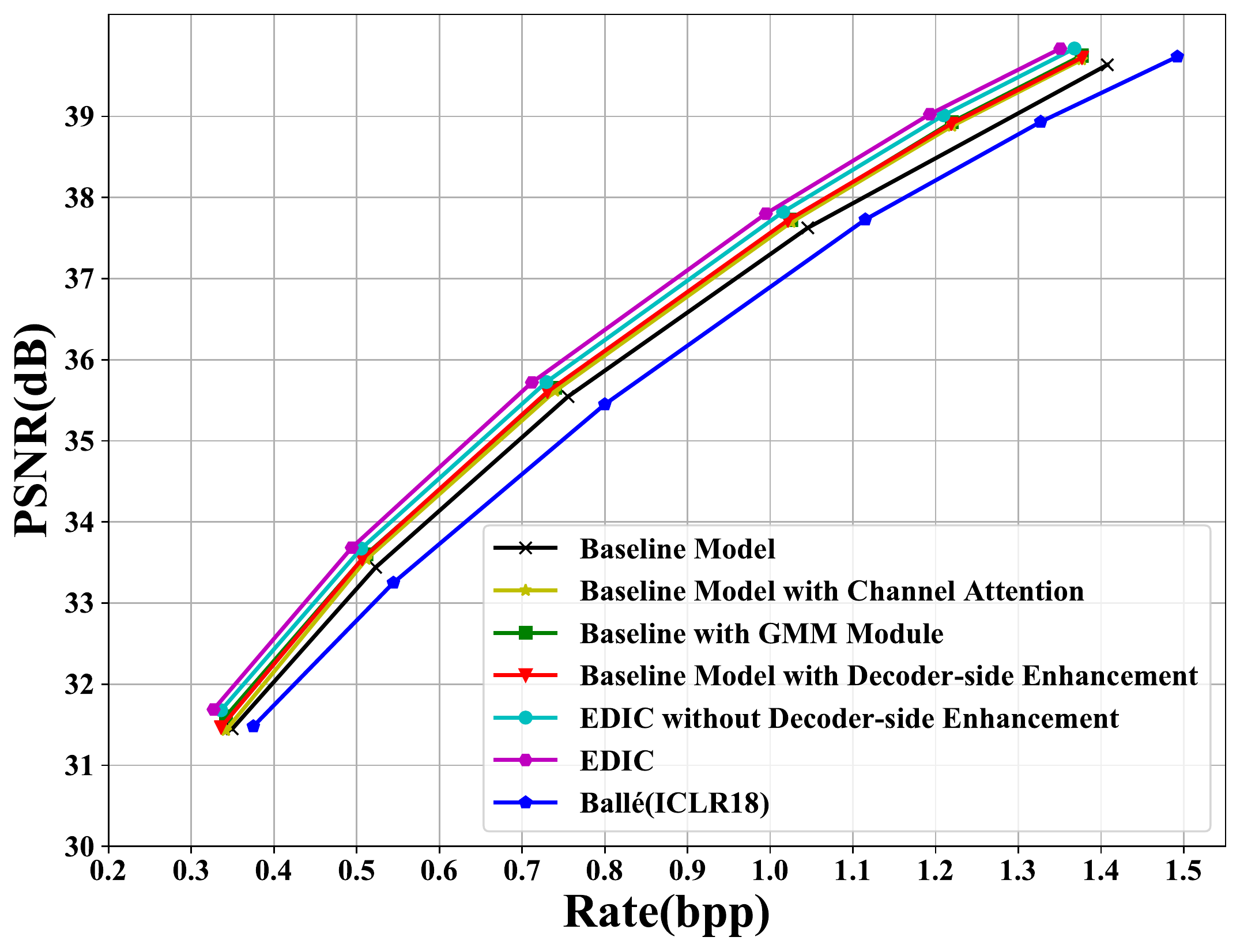}
  \caption{
     Effectiveness of each module in our newly proposed framework.
  }
  \label{figure: ablation}
\end{figure}
\begin{figure*}[htp]

  % \subfigure[Rate-distortion curves of our method and competetive methods based on MS-SSIM quality metric.]{
    \centering
  \begin{minipage}[b]{0.85\textwidth}
    \includegraphics[width=0.5\textwidth]{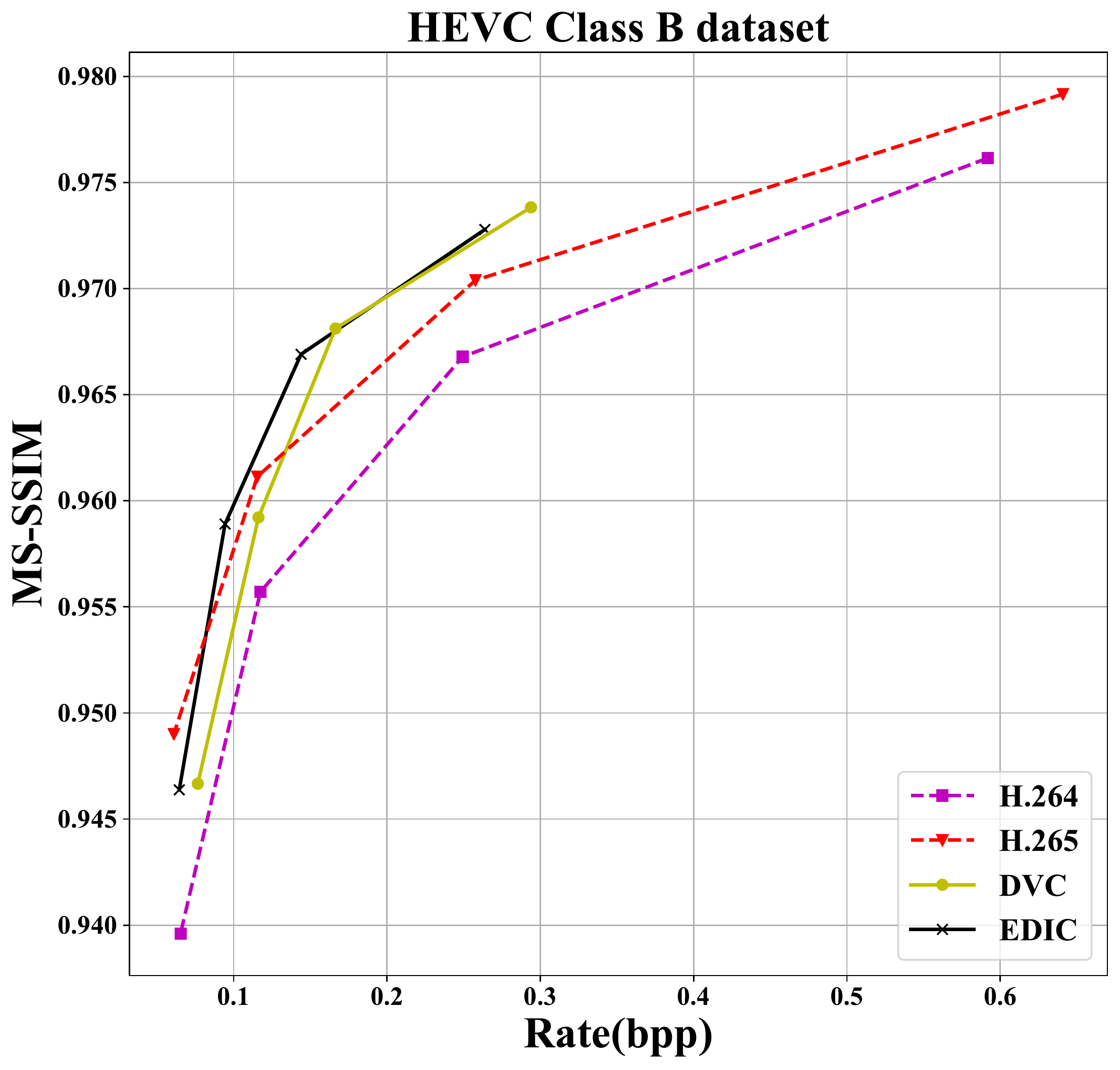}
    \includegraphics[width=0.5\textwidth]{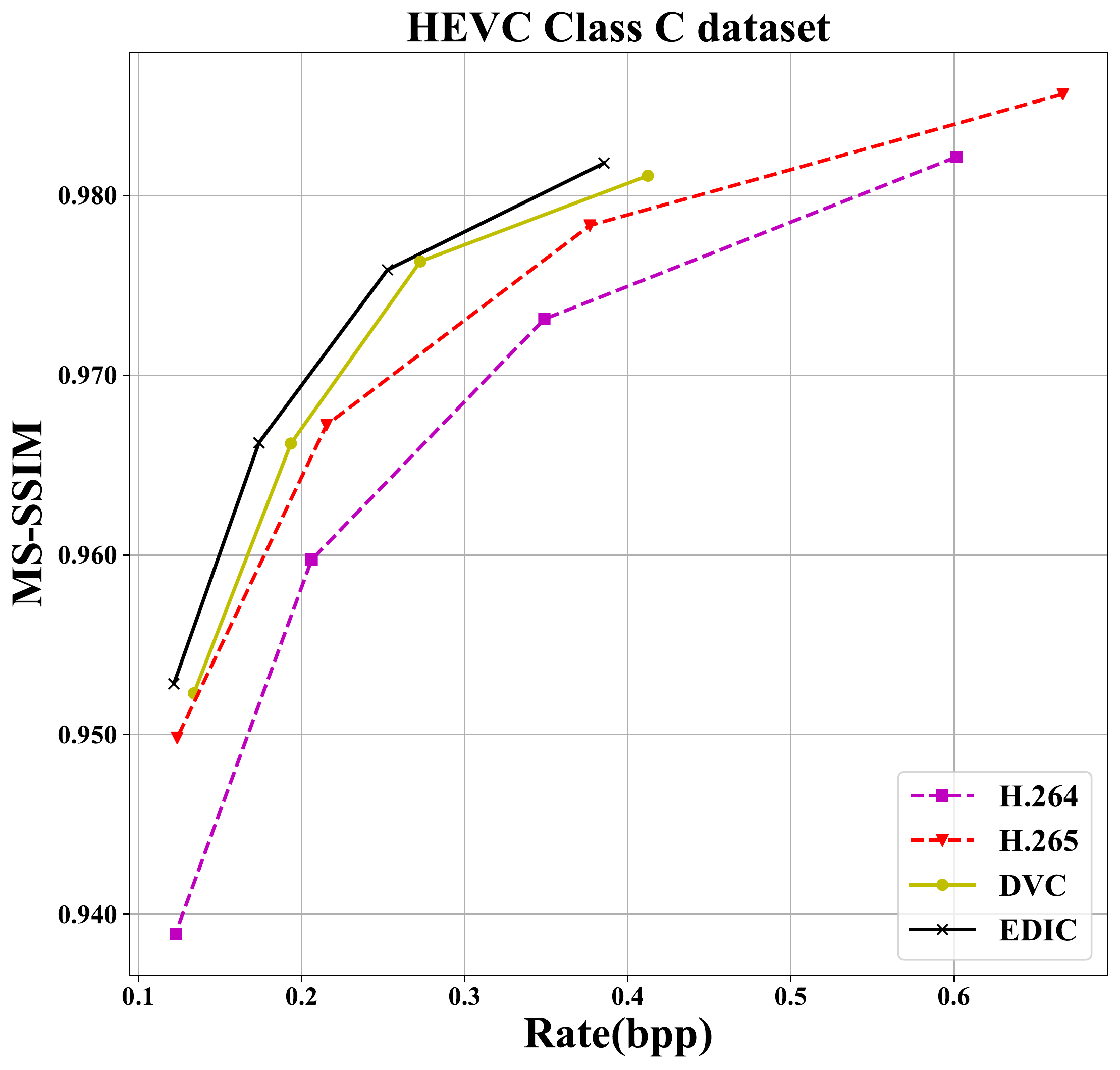}
    \includegraphics[width=0.5\textwidth]{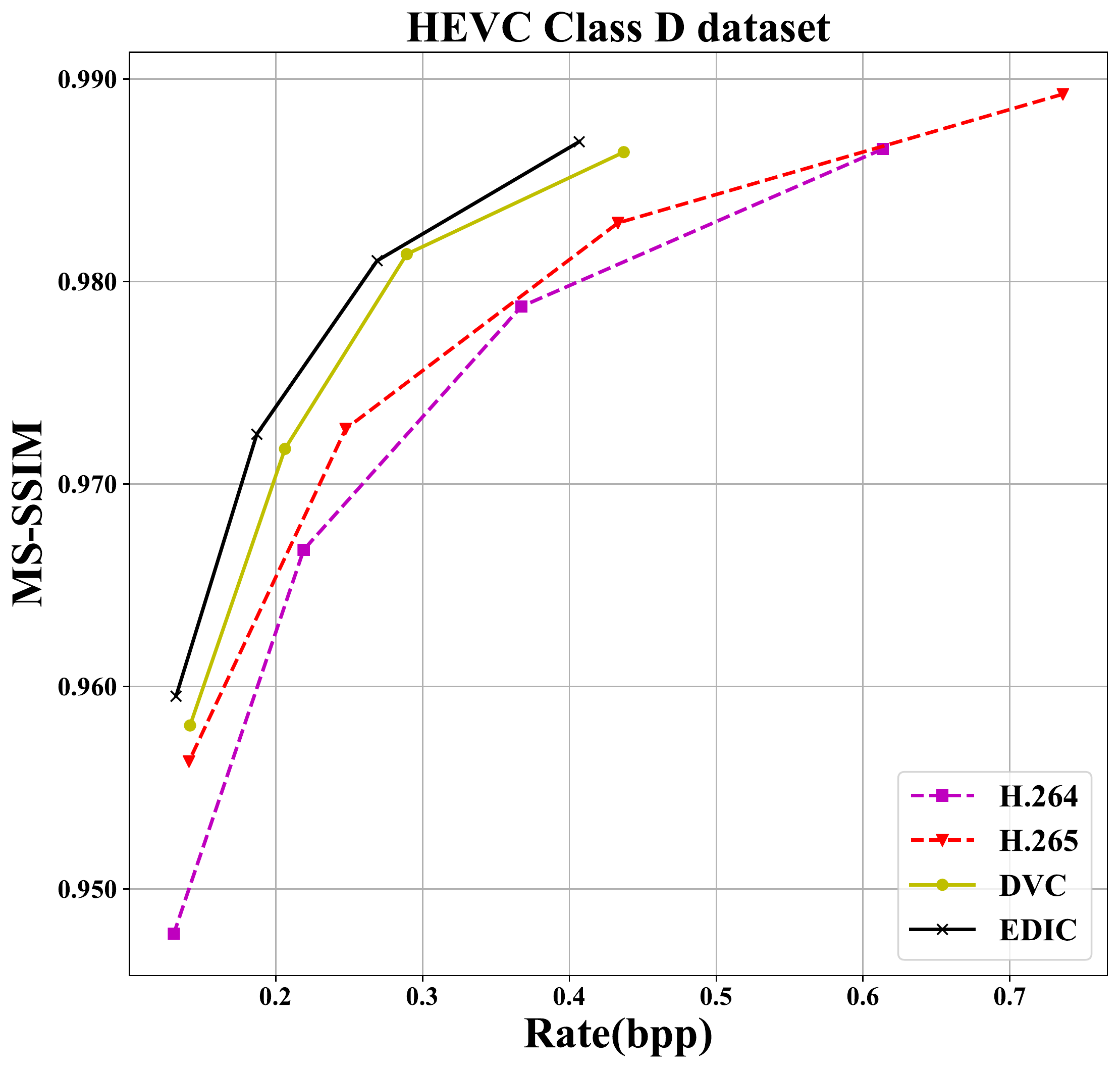}
    \includegraphics[width=0.5\textwidth]{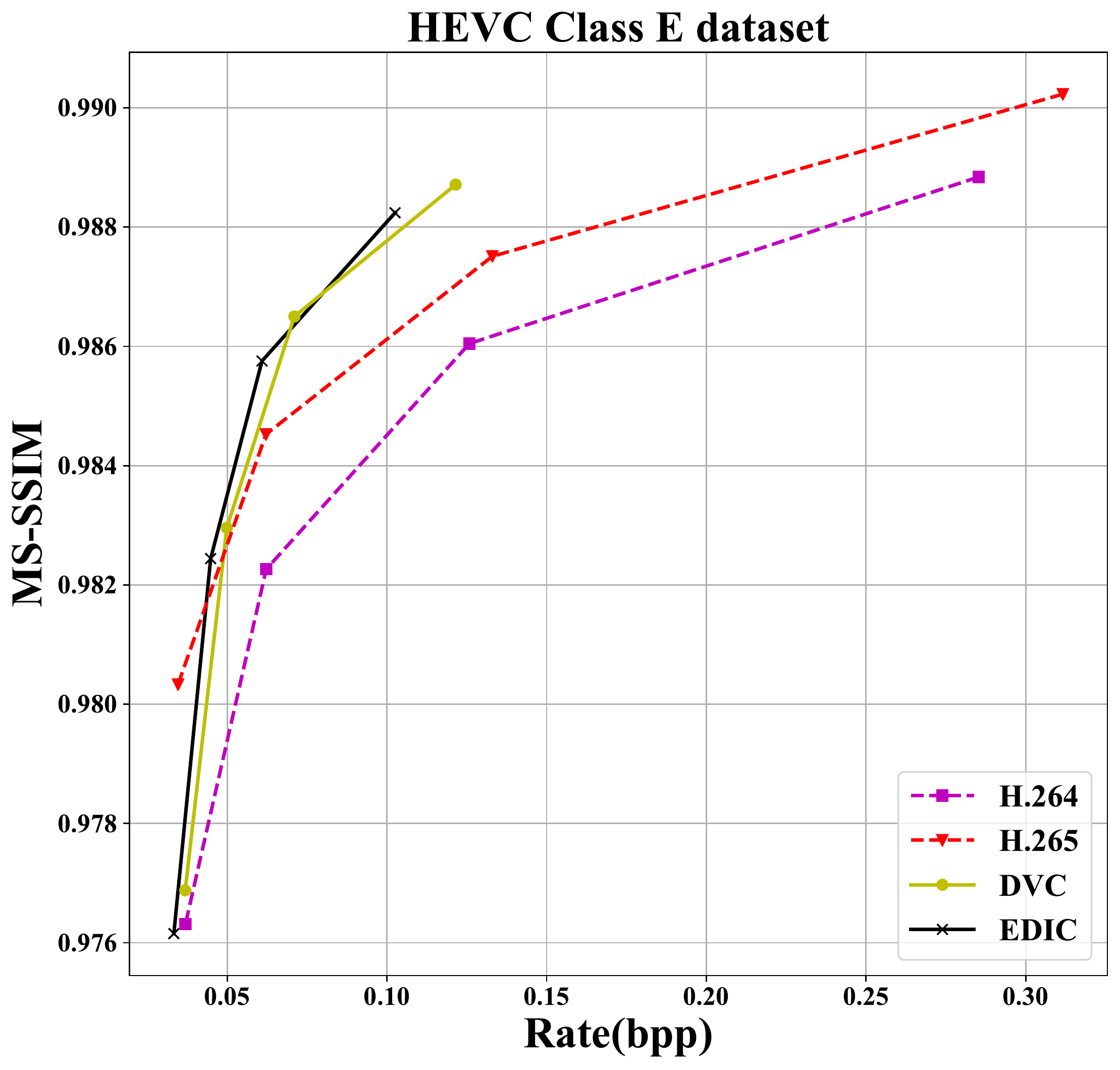}  
  \end{minipage}
  \caption{Rate-distortion curves of our proposed EDIC method and the competitive methods for video compression when using the MS-SSIM metric.}
  \label{fig:subfig}
\end{figure*}
\begin{figure}[htp]
\centering
\includegraphics[width=1.0\linewidth]{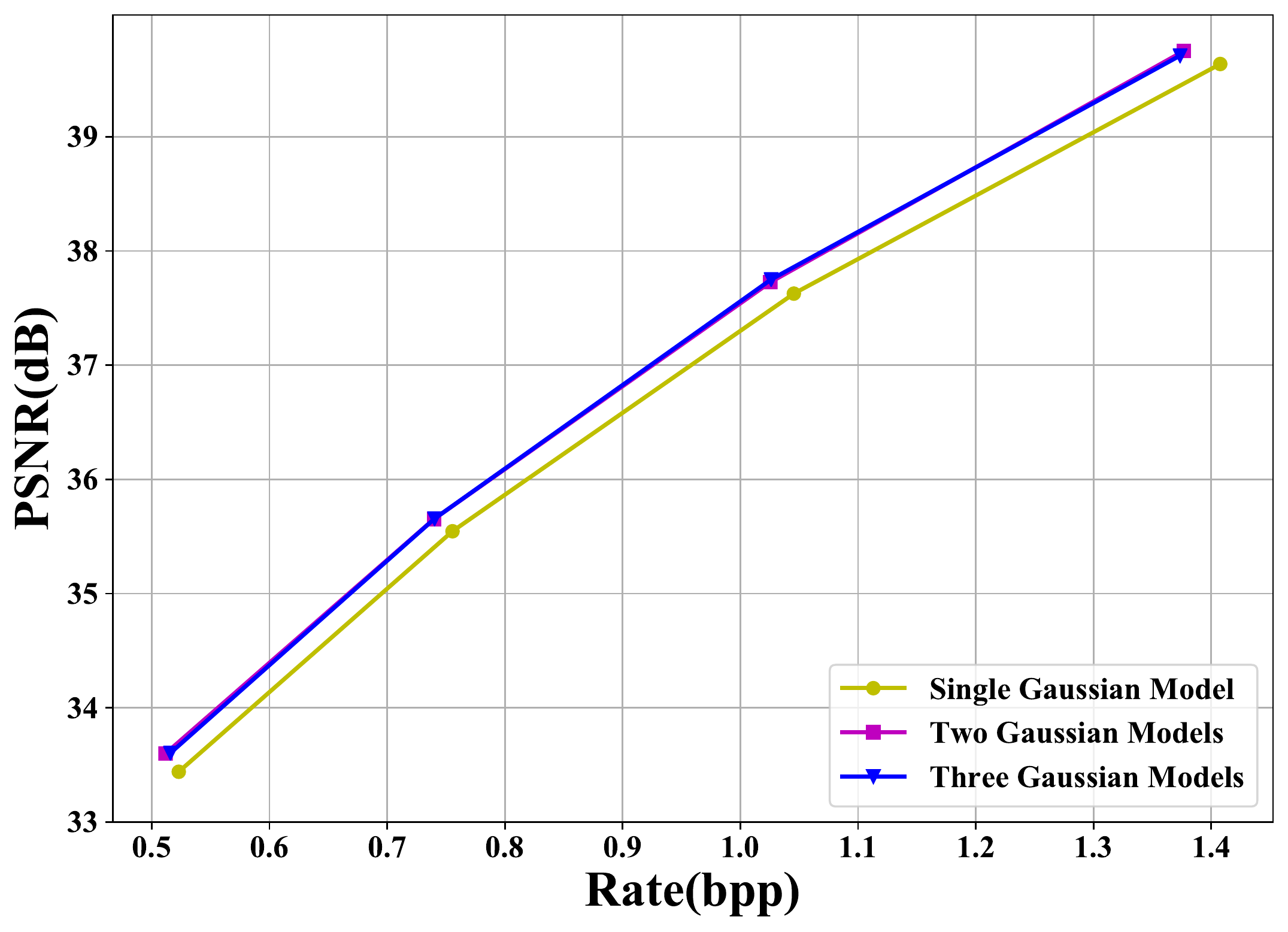}
\caption{
   The results when using different numbers of Gaussian models in our method.
}
\label{figure: number}
\end{figure}
\subsection{Performance and implementation details for Video Compression}
For the implementation details on video compression,
we follow the settings of DVC~\cite{lu2019dvc}.
Each video clip in the Vimeo dataset consists of 7 frames. 
% For each frame in the video clip, we use $X_t$ as a reference frame and $X_{t+2}$ as the original frame during the training process.
The HEVC test dataset contains the videos with different resolutions and different contents.
We set $N$ to 192 and $M$ to 288 in Fig.~\ref{figure: architecture}.
Other modules of our video compression framework is same as DVC~\cite{lu2019dvc}.
In the training process, when using the quality metric as the MSE loss function,
we select $\lambda={4096}$ to obtain our pre-trained model for 2,000,000 iterations with the learning rate of ${1\times10^{-4}}$.
Then, we apply different $\lambda$ values (\textit{i.e.}, 256, 512, 1024, 2048) to
fine-tune this pre-trained model with the learning rate of ${1\times10^{-5}}$.
When optimized by using the MS-SSIM loss function, we fine-tune the model at high bitrates from the MSE loss function for 80,000 iterations with the learning rate of ${1\times10^{-5}}$.
The remaining training strategies are similar to the implementation details of image compression described in Section~\ref{details}.

As shown in Fig.~\ref{fig:subfig}, we compare our method with the traditional video compression standards, like H.264~\cite{x264}, H.265~\cite{x265},
and the deep learning based method DVC~\cite{lu2019dvc}.
To obtain the compressed frames by the H.264 and H.265, 
we apply the FFmpeg with very fast mode, and  set the GOP sizes of the HEVC dataset to 10.
As for the Rate-Distortion(RD) curves,
in terms of the PSNR quality metric, our method is much better than DVC~\cite{lu2019dvc} and H.264~\cite{x264},
and it achieves comparable performance
with H.265~\cite{x265}.
With regard to MS-SSIM,
it is clear that our newly proposed method is superior to DVC~\cite{lu2019dvc}, H.264~\cite{x264}, and H.265~\cite{x265} for almost all the HEVC test classes.

% In each row, the left-most column denotes the result of JPEG~\cite{wallace1992jpeg} compression method, the middle left column refers to the BPG~\cite{BPG},
% the middle column is Ball{\'{e}}'s method~\cite{balle2018variational}.
% the middle right column represents  and the right-most column is our proposed E-DIC method.}
\begin{figure*}[htp]
  \centering
  \includegraphics[width=1.0\linewidth]{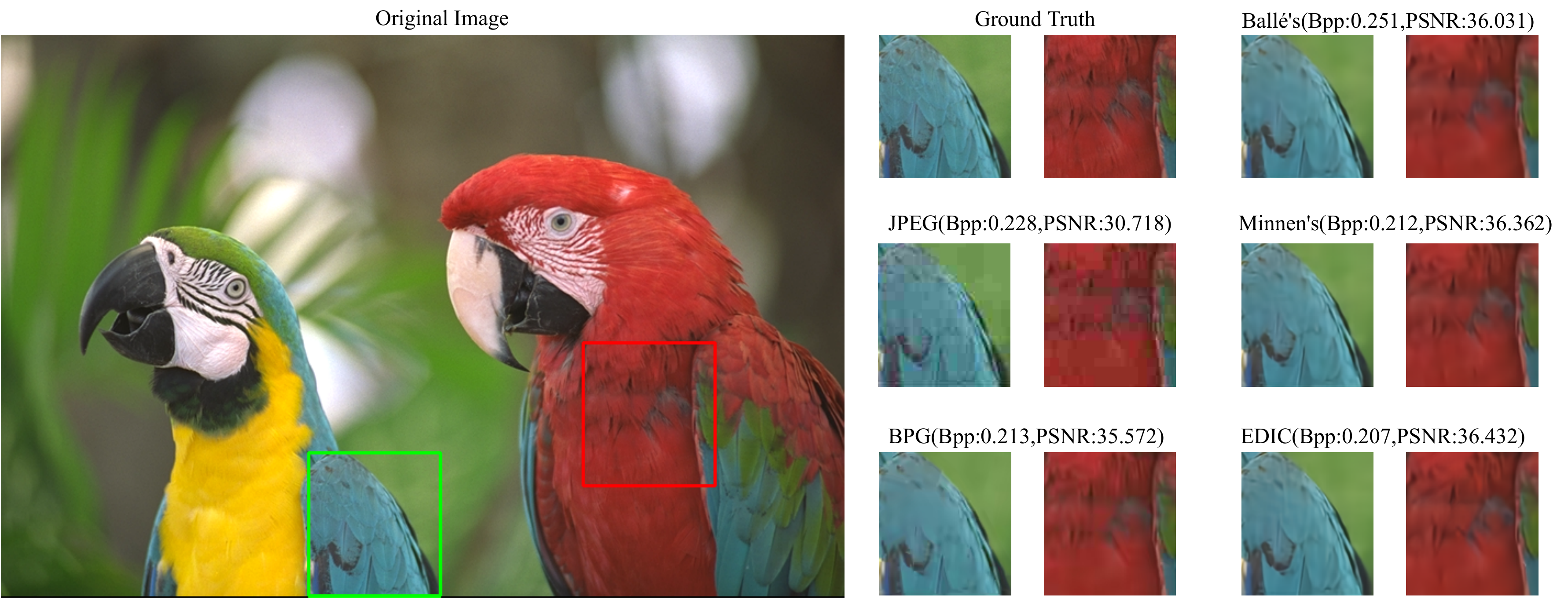}
  \caption{
    Visualization of reconstructed sample images of ground truth, JPEG~\cite{wallace1992jpeg}, BPG~\cite{BPG},
    Ball{\'{e}}'s method~\cite{balle2018variational}, Minnen's method~\cite{minnen2018joint} and our proposed EDIC method.
    We take ``kodim23.png'' from Kodak~\cite{Kodak} for illustration.
  }
  \label{figure: sample}
\end{figure*}
\subsection{Ablation study}
\subsubsection{Effiectiveness of Each module}
In order to verify the effectiveness of each proposed module, we perform ablation study for image compression in this section.
For the baseline model, we utilize a single Gaussian model as our entropy model.
When implementing the baseline model,
we just remove the GMM module, the attention module and the decoder-side enhancement module (See Fig.~\ref{figure: architecture}).
After that, the last convolution layer of the hyper decoder is $2*N$, so the first $N$ channels are used to estimate the mean parameters and the last $N$ channels are used to estimate the 
variance parameters of a single Gaussian model.
Then, we add each module to the baseline model, respectively.
When we utilize the Gaussian mixture model as our entropy model,
we add the GMM module described in Fig.~\ref{fig:gmm} based on the baseline model to estimate the parameters of the Gaussian mixture model.
As shown in Fig.~\ref{figure: ablation}, we compare the performance of the baseline model,
the baseline model with additional attention module,
the baseline model with additional decoder-side enhancement module,
the baseline model with additional GMM module,
our proposed EDIC method without decoder-side enhancement module,
and our overall EDIC method consisting of the GMM module, the attention module and the decoder-side enhancement module.
% We also add the results of ~\cite{minnen2018joint} to demonstrate the effectiveness of our method clearly.
For all experiments, we use the same training strategy described in Section~\ref{details}.
As shown in Fig.~\ref{figure: ablation},
we observe that each module brings significant performance improvement when compared to our baseline model.
For the attention module, the baseline model with the attention module is about 0.2 dB better than to the baseline model.
The baseline model with the GMM module is also superior to the baseline model, which demonstrates the effectiveness of the Gaussian mixture model.
Furthermore, when we add the decoder-side enhancement module to any models, we can achieve better performance.
\subsubsection{Results when using different numbers of Gaussian Models}
We also conduct the experiments to report the results when using different numbers of Gaussian models for image compression.
Specifically, we adopt two Gaussian models in our implementation of the Gaussian mixture model.
When we use three Gaussian models, 
we simply change the number of output channels in the last layer of GMM module (See Fig.~\ref{fig:gmm}) to
$9\times{N}$. The first $6\times{N}$ channels estimate the parameters of mean and variance,
while the last $3\times{N}$ channels estimate the weights of each Gaussian model.
In order to make the sum of weights equal to 1,
we add the softmax layer to the output of the last $3\times{N}$ channels.
As shown in Fig.~\ref{figure: number},
we observe that our method using three Gaussian models achieve similar performance with that using two Gaussian models,
which demonstrates that the performance of our approach cannot be improved significantly when increasing the number of Gaussian models.
\begin{figure}[htp]
  \centering
  % \subfigure[Bit allocation map.]{
      % \centering
  \begin{minipage}[b]{0.42\textwidth}
    \centering
    \includegraphics[width=0.48\textwidth]{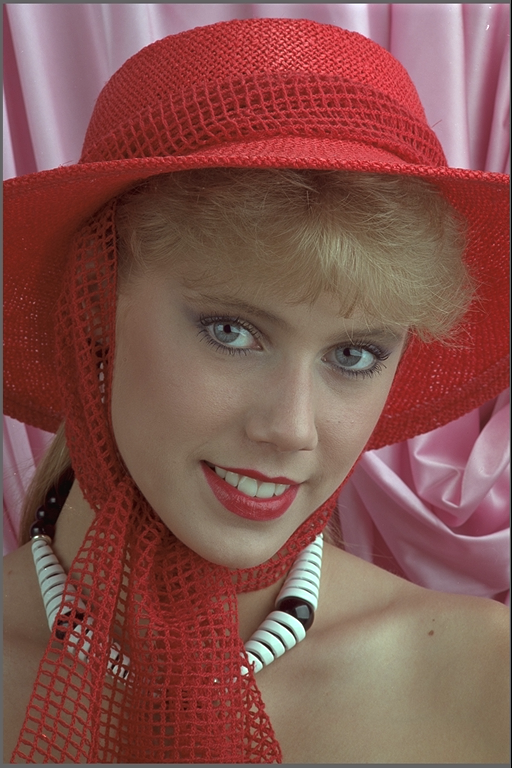}
    \includegraphics[width=0.48\textwidth]{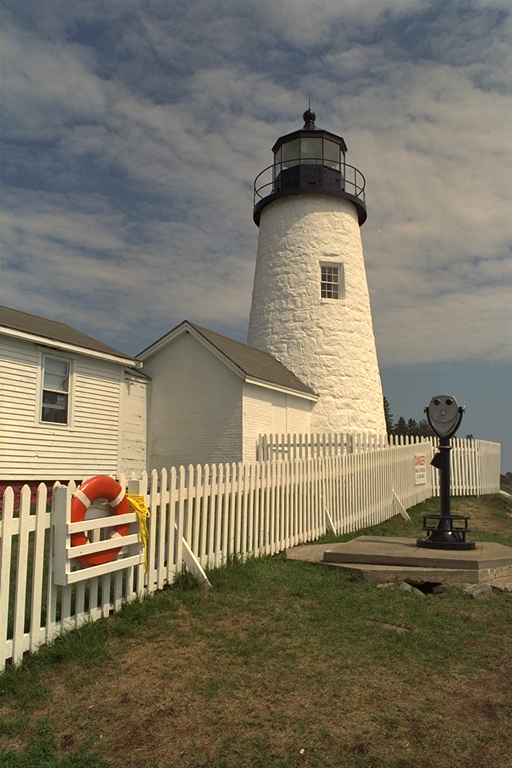}
  \end{minipage}
  \\[2pt]
  % \subfigure[We take ``kodim04.png'', and ``kodim19.png'' from ~\cite{Kodak}.]{
      \centering
  \begin{minipage}[b]{0.42\textwidth}
    \centering
    \includegraphics[width=0.48\textwidth]{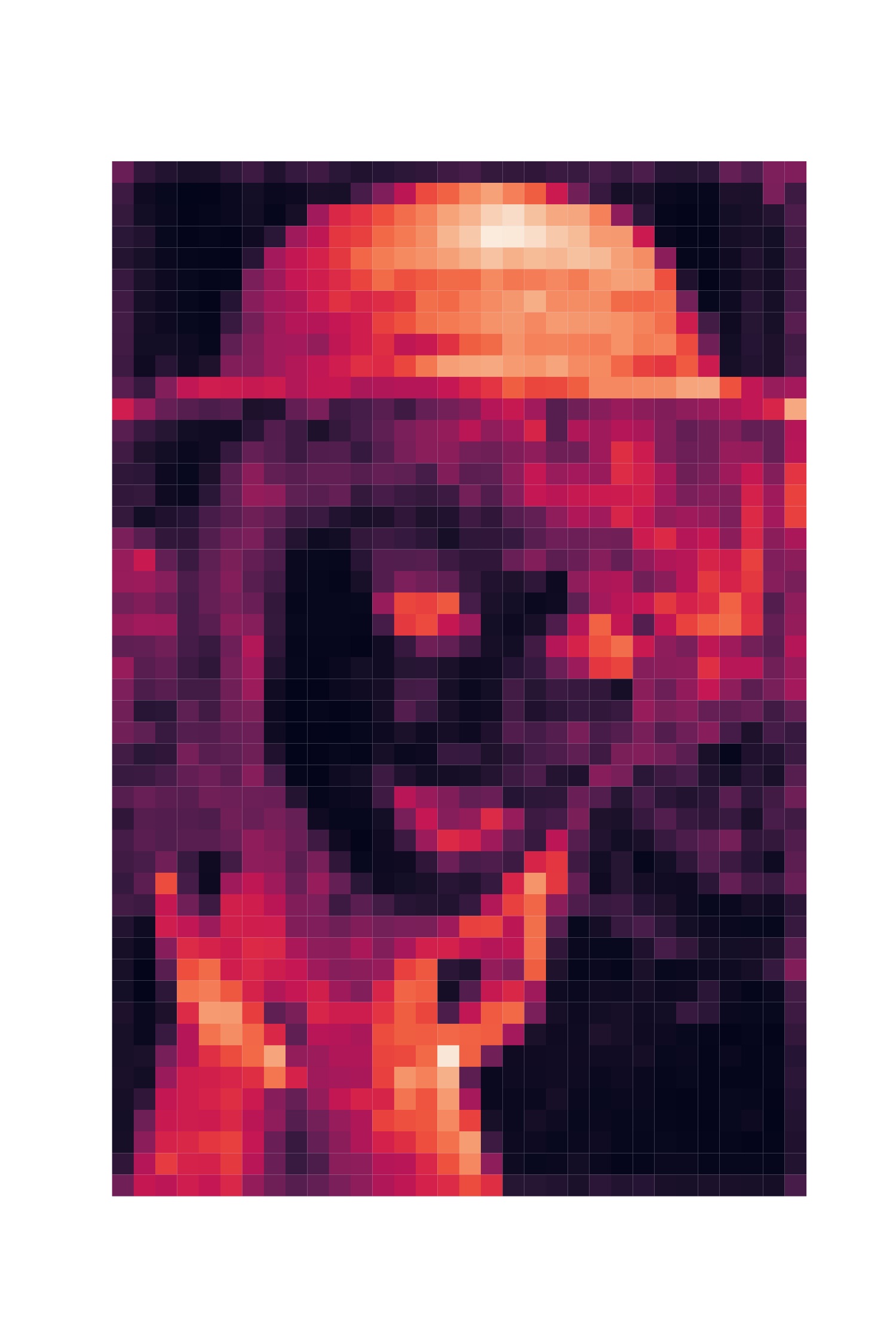}
    \includegraphics[width=0.48\textwidth]{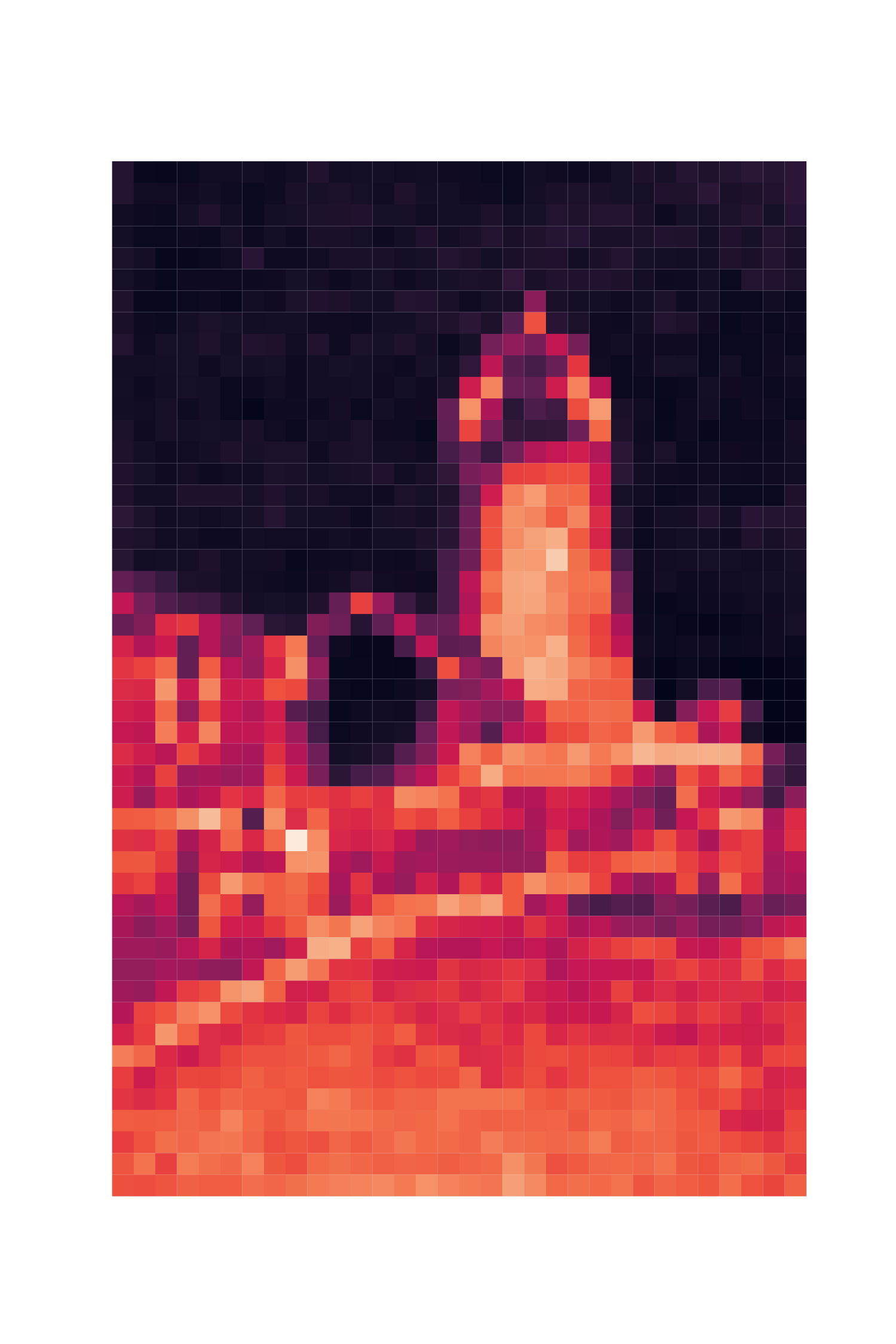} 
  \end{minipage}
  % }
  \caption{Comparsion between the bit allocation map of latent representations $y$ (the second row) and original image (the first row).
  We take ``kodim04.png'', and ``kodim19.png'' from Kodak~\cite{Kodak} for illustration.}
  \label{fig:bitmap}
\end{figure}
\subsection{Visualization}
In order to demonstrate the effectiveness of our EDIC more clearly,
we provide some visualization results.
As shown in Fig.~\ref{fig:bitmap}, we visualize the bit allocation map of latent representations $\hat{y}$.
The brighter region means we allocate more bits.
In the smooth region, our proposed EDIC method allocates a few bits.
In contrast, we need more bits in the edge region, 
which means that our neural network can learn to allocate bits according to different types of regions automatically.
Furthermore, we compare the reconstructed sample images of our proposed EDIC method and other competitive methods in Fig.~\ref{figure: sample}.
The results of the learned image compression methods are optimized by MSE loss function.
The reconstructed image of our method achieves higher quality in both PSNR metric and qualitative viewing when the compression ratios of all methods are close.

\section{Conclusion}
In this paper, we have proposed a unified framework EDIC to boost image compression performance while keeping fast inference speed for practical scenarios.
We first adopt a light-weight channel-wise attention mechanism to reduce channel-wise redundancy of the latent representations. Moreover, we propose to use the Gaussian mixture model to estimate the
bitrate more accurately, which has been shown to be very useful for edge regions.
Finally, we introduce a simple decoder-side enhancement module to further improve image compression performance.
Our framework can be trained in an end-to-end fashion and readily used for video compression.
Experimental results have demonstrated the superiority of our proposed EDIC method for image and video compression over the existing state-of-the-art methods.

% or


% \appendices

% % you can choose not to have a title for an appendix
% % if you want by leaving the argument blank
% \section{}

% % use section* for acknowledgment
% \section*{Acknowledgment}


\ifCLASSOPTIONcaptionsoff
  \newpage
\fi

\bibliographystyle{IEEEtran}
\bibliography{egbib}
\end{document}